\documentclass[12pt]{article}
\usepackage{amssymb,makeidx}
\usepackage[dvips]{graphicx}
\usepackage{cite}
\newcommand{\be}{\begin{eqnarray}}
\newcommand{\ee}{\end{eqnarray}}
\newcommand{\bez}{\begin{eqnarray*}}
\newcommand{\eez}{\end{eqnarray*}}
\newcommand{\pa}{\partial}
\newcommand{\na}{\nabla}
\renewcommand{\d}{{\rm d}}
\newcommand{\dl}{\delta}
\newcommand{\ad}{{\rm ad}}
\newcommand{\A}{{\cal A}}
\newcommand{\X}{{\cal X}}
\newcommand{\g}{\mathfrak{g}}
\newcommand{\h}{\mathfrak{h}}
\newcommand{\oA}{\otimes_\A}

\newcommand{\V}{{\cal V}}
\newcommand{\R}{{\cal R}}
\newcommand{\oL}{\otimes_L}

\newcommand{\hmu}{{\hat{\mu}}}
\newcommand{\hnu}{{\hat{\nu}}}
\newcommand{\hrho}{{\hat{\rho}}}

\newcommand{\hi}{{\hat{\imath}}}
\newcommand{\hj}{{\hat{\jmath}}}
\makeindex

\title{\bf Riemannian Geometry \\
           of Bicovariant Group Lattices}
\date{  }
\author{{\bf Aristophanes Dimakis}\thanks{Electronic mail: dimakis@aegean.gr}  \\
 Department of Financial and Management Engineering, \\
 University of the Aegean, 31 Fostini Str., GR-82100 Chios
 \and
 {\bf Folkert M\"uller-Hoissen}\thanks{Electronic mail: fmuelle@gwdg.de}\\
 Max-Planck-Institut f\"ur Str\"omungsforschung, \\
 Bunsenstrasse 10, D-37073 G\"ottingen
}

\begin{document}

\maketitle

\renewcommand{\theequation} {\arabic{section}.\arabic{equation}}

\newtheorem{theorem}{Theorem}[section]
\newtheorem{lemma}{Lemma}[section]

\newcounter{example}[section]
\renewcommand{\theexample}{\arabic{section}.\arabic{example}}
\newenvironment{example}
{ \refstepcounter{example} \noindent
  {\it Example \arabic{section}.\arabic{example}}.}
{ \hfill $\blacksquare$ \vspace{.2cm} }

\vskip1cm

\begin{abstract}
Group lattices (Cayley digraphs) of a discrete group are in natural
correspondence with differential calculi on the group. On such a
differential calculus geometric structures can be introduced following
general recipes of noncommutative differential geometry.
Despite of the non-commutativity between functions and
(generalized) differential forms, for the subclass of ``bicovariant''
group lattices considered in this work it is possible to understand
central geometric objects like metric, torsion and curvature as
``tensors'' with (left) covariance properties. This ensures
that tensor components (with respect to a basis of the space
of 1-forms) transform in the familiar homogeneous way under
a change of basis. There is a natural compatibility condition
for a metric and a linear connection. The resulting (pseudo-)
Riemannian geometry is explored in this work. It is demonstrated
that the components of the metric are indeed able to properly
describe properties of discrete geometries like lengths and angles.
A simple geometric understanding in particular of torsion and curvature
is achieved. The formalism has much in common with
lattice gauge theory. For example, the Riemannian curvature is
determined by parallel transport of vectors around a plaquette
(which corresponds to a biangle, a triangle or a quadrangle).
\end{abstract}

\newpage

\small
\tableofcontents
\normalsize

\newpage

\section{Introduction}
\setcounter{equation}{0}
In a previous paper \cite{DMH02-gl} we started to develop a general
formalism of differential geometry of group lattices (Cayley digraphs),
based on elementary notions of noncommutative geometry.
The present work extends the latter to a formalism of discrete
(pseudo-) Riemannian geometry of the subclass of \emph{bicovariant} group
lattices, as defined in Ref.~\citen{DMH02-gl}.
A group lattice, which is determined by a \emph{discrete} group
$G$ and a \emph{finite} subset $S$
(which does not contain the unit element $e$) naturally defines a
first-order differential calculus (which extends to higher orders)
over the algebra $\A$ of functions on $G$.
If $S$ generates $G$, then bicovariance of the group lattice $(G,S)$
is equivalent to bicovariance of the first-order differential calculus
in the sense of Ref.~\citen{Woro89}.
\vskip.1cm

``Riemannian geometry'' of discrete groups in the context of
noncommutative geometry has already been considered in
several publications \cite{BDMHS96,Cast,Majid}. The present approach differs
from these in particular by introducing a metric tensor as an element of
a left-covariant tensor product of the space of 1-forms with itself.
This tensor product is obtained from the a priori naturally given
tensor product over $\A$ by using the special structure of group
lattices and the bicovariance condition. Although
this formalism has some ideas in common with the approach of
Ref.~\citen{DMH99-dRg}, it crucially differs from the latter, where a
left-covariant tensor product for arbitrary differential calculi on finite sets
was constructed making use of a connection. The present approach is much simpler
and geometrically more transparent, but restricted to bicovariant group
lattices and thus a subclass of \emph{regular} \cite{reg_graphs} digraphs.
One should keep in mind that extensions of geometric structures from
ordinary differential geometry to the framework of noncommutative geometry may
be carried out in various ways and only applications can decide on their
usefulness. For our choice, we will demonstrate that it leads to simple and
convenient rules of discrete geometry. It is also this last aspect, namely
the fact that we establish a geometric interpretation of the a priori
abstract formalism, which distinguishes the present work from some
previous publications on noncommutative geometry of discrete groups.
\vskip.1cm

The reason why we define a metric as a left-covariant tensor is that
in this case its components are ``local'' objects (see
section~\ref{subsec:nonlocal} for details).
More generally, the components of left-covariant tensors obey a
homogeneous local transformation law under a change of basis.
In this sense they are really counterparts of tensors in
ordinary differential geometry. This is quite in the spirit
of Wilson's lattice gauge theory: discretization a priori
moves local fields to non-local objects, but via parallel
transport around a plaquette local objects are obtained. This
is important in order to maintain gauge invariance, which is
the main principle behind it. Similarly, we may postulate
the preservation of the tensor transformation principle.
This also allows to consider coordinate transformations on
group lattices very much in analogy with continuum differential
geometry (see section~\ref{sec:coord}). The idea of constructing
left- or alternatively right-covariant tensors in a noncommutative
differential calculus already appeared in Ref.~\citen{DMH92}.
Viewed as a map between left $\A$-modules, a left-covariant
tensor is left $\A$-linear.
\vskip.1cm

Discrete (pseudo-) Riemannian geometry is of relevance
for numerical evaluation and also path integral quantization of
classical physical models based on continuum Riemannian geometry,
like mechanical and general relativistic systems (see Ref.~\citen{DGP02},
for example). The approach based on concepts of noncommutative
geometry is an alternative to Regge calculus \cite{Regge61}. It has the advantage,
however, that its formal structure is much closer to continuum
differential geometry. Similarities with previous approaches
to gravity using concepts of lattice gauge theory exist \cite{latt_grav},
but there is little overlap when it comes to the details of the
formalism. Of course, discrete geometry is an old subject (see Ref.~\citen{Sauer70},
for example) and relations between the present work and earlier
approaches can certainly be established to some extent. This will not be
attempted in this work. Rather, we concentrate on what the machinery
of algebraic noncommutative geometry applied in a natural way
to (bicovariant) group lattices gives us and we reveal the geometric
significance of discrete analogues of metric, metric-compatible
linear connections, torsion and curvature.
\vskip.1cm

Section~\ref{sec:tensor} discusses the non-locality of the tensor product over
$\A$ and introduces the left-covariant tensor product for bicovariant group
lattices, which induces a left-covariant product in the space of forms $\Omega$.
Left-covariant metrics are then introduced and a compatibility condition
with a linear connection is formulated. The geometric meaning of the
parallel transport determined by a metric-compatible linear connection
is explored. Furthermore, we introduce the notion of a ``discrete
Killing vector field''.
\vskip.1cm

Section~\ref{sec:tor&curv} elaborates the torsion and the curvature
of linear connections on bicovariant group lattices and also
provides corresponding expressions in terms of basic vector fields
(which constitute a subclass of discrete vector fields, see
Ref.~\citen{DMH02-gl}). Appendix A presents expressions of basic
formulas with respect to an orthonormal coframe field.
\vskip.1cm

Section~\ref{sec:discrRiem} deals with group lattices which carry
a metric and a torsion-free compatible linear
connection. Several examples are treated. The conclusion is that
for most group lattices a restriction to torsion-free metric-compatible
linear connections too severely restricts the possible geometries.
\vskip.1cm

A metric-compatible linear connection provides us with a parallel
transport which maps part of the group lattice isometrically into
the tangent space at some site. Torsion and curvature of the
connection are, respectively, corresponding first and second order
obstructions. As we demonstrate with several examples, in particular
in section~\ref{sec:gl-torsion}, torsion plays
a much more fundamental role in this discrete framework than in
ordinary continuum differential geometry. Linear connections with
torsion are needed to describe even simple group lattice geometries
in this framework.
\vskip.1cm

In section~\ref{sec:coord} we introduce the concept of coordinates
on group lattices and elaborate in particular the geometry of
hypercubic lattices based on the Abelian group $\mathbb{Z}^n$.
Some concluding remarks are collected in section~\ref{sec:concl}.
\vskip.1cm

The present work relies on the notation and results of
Ref.~\citen{DMH02-gl}. It is \emph{not} self-contained. We refer to
an equation in Ref.~\citen{DMH02-gl} in the form $(I.a.b)$ where
$(a.b)$ is the equation number in Ref.~\citen{DMH02-gl}.
In the following we restrict our considerations to bicovariant group
lattices $(G,S)$. This means that $S$ is assumed to be closed under
the adjoint action of all elements of $S$ and their inverses.

\section{Tensor products, metrics, and linear connections}
\label{sec:tensor}
\setcounter{equation}{0}
In this section we first briefly discuss the consequences of the
non-locality of the usual tensor product over $\A$. Then we make
use of the special structure of bicovariant group lattices in
order to construct a new tensor product which is left-covariant
so that the corresponding tensor components are ``local'' and
able to carry a geometric meaning. The left-covariant tensor product
of forms induces a left-covariant (generalized wedge) product
in the space of forms. Left-covariant metrics are introduced
and a compatibility condition with a linear
connection is formulated. The latter involves an extension of the
linear connection from the space of 1-forms to a left-covariant
tensor product. This is a familiar procedure in the tensor calculus
on manifolds, but in general not at all straight forward in
noncommutative geometries (see also Ref.~\citen{BDMHS96}).
Of particular importance for an understanding of the formalism
is the observation that a metric-compatible
linear connection determines an isometric map of parts of
the group lattice into the tangent space at a (fixed) site.
In the last subsection we define discrete Killing vector fields
and invariant metrics on a (bicovariant) group lattice.

\subsection{Non-locality of the tensor product over $\A$}
\label{subsec:nonlocal}
For the differential calculus $(\Omega,\d)$ determined by a group lattice
$(G,S)$ there is a distinguished (left and right) $\A$-module basis
$\{ \theta^h \, | \, h \in S \}$ of the space of 1-forms $\Omega^1$
which satisfies $\theta^h f = R^\ast_h f \, \theta^h$
for all elements $f$ of the space of functions $\A$ on $G$,
where $R_h$ is the right action on $G$ by an element $h \in S$.
As a consequence,
\be
   (f \theta^h) \oA (f' \theta^{h'})
 = f \, (R^\ast_h f') \, \theta^h \oA \theta^{h'}
\ee
for all $f,f' \in \A$. For each $g \in G$ there is a function
$e^g$ such that $e^g(g') = \delta_{g,g'}$ for all $g' \in G$.
For this function we obtain
\be
   e^g (\theta^h \oA \theta^{h'}) = (e^g \theta^h) \oA (e^{gh} \theta^{h'})
           \label{e^g-oA}
\ee
which shows that the tensor product $\oA$ is non-local since the two
factors ``sit'' at different (though neighboring) points.
Let us consider an object
\be
  \mathsf{g} = \sum_{h,h' \in S} \gamma_{h,h'} \, \theta^h \oA \theta^{h'}
           \label{g_oA}
\ee
with $\gamma_{h,h'} \in \A$. Under a linear change of basis
$\theta^h \mapsto \tilde{\theta}^h := \sum_{h' \in S} a^h{}_{h'} \, \theta^{h'}$
with coefficients $a^h{}_{h'} \in \A$ we find
\be
   \mathsf{g}
 = \sum_{h_1,h_2,h'_1,h'_2 \in S} \gamma_{h'_1,h'_2} \,
   (a^{-1})^{h'_1}{}_{h_1} \, (R^\ast_{h_1} a^{-1})^{h'_2}{}_{h_2} \,
   \tilde{\theta}^{h_1} \oA \tilde{\theta}^{h_2}
 = \sum_{h_1,h_2 \in S} \tilde{\gamma}_{h_1,h_2} \,
   \tilde{\theta}^{h_1} \oA \tilde{\theta}^{h_2}
\ee
from which we read off the coefficients with respect to the new cobasis:
\be
   \tilde{\gamma}_{h_1,h_2}
 = \sum_{h'_1,h'_2 \in S} \gamma_{h'_1,h'_2} \,
   (a^{-1})^{h'_1}{}_{h_1} \, (R^\ast_{h_1} a^{-1})^{h'_2}{}_{h_2}  \; .
\ee
Here we see again the non-local character of the tensor product $\oA$.

\subsection{Left-covariant tensor product for bicovariant group lattices}
By acting on each component, the maps
$R_h^\ast$ and $R_{h^{-1}}^\ast$ for $h \in S$ extend to tensor
products of $\Omega^1$ and to $\Omega$ as automorphisms.
Then there is another tensor product with a local transformation rule.
This ``left-covariant'' tensor product is defined via
\be
    (\theta^{h_1} \oA \ldots \oA \theta^{h_r}) \oL T
 := \theta^{h_1} \oA \ldots \oA \theta^{h_r} \oA R^\ast_{h_r^{-1}} \cdots
    R^\ast_{h_1^{-1}} T          \label{oL_def1}
\ee
where $T$ is an arbitrary element of a tensor product of $\Omega^1$ over $\A$.
The inverse relation is
\be
    (\theta^{h_1} \oA \ldots \oA \theta^{h_r}) \oA T
  = (\theta^{h_1} \oA \ldots \oA \theta^{h_r}) \oL R^\ast_{h_1} \cdots
    R^\ast_{h_r} T    \; .
\ee
Using $R_h^\ast \, \theta^{h'} = \theta^{\ad(h)h'}$ we find in particular
\be
    \theta^h \oL \theta^{h'}
  = \theta^h \oA \theta^{\ad(h^{-1}) h'} \, , \qquad
    \theta^h \oA \theta^{h'}
  = \theta^h \oL \theta^{\ad(h) h'}  \; .
\ee
Note also that
\be
    (\theta^{h_1} \oL \ldots \oL \theta^{h_r}) \oL T
  &=& (\theta^{h_1} \oL \ldots \oL \theta^{h_r}) \oA R^\ast_{h_1^{-1}} \cdots
    R^\ast_{h_r^{-1}} T   \\
    (\theta^{h_1} \oL \ldots \oL \theta^{h_r}) \oA T
  &=& (\theta^{h_1} \oL \ldots \oL \theta^{h_r}) \oL R^\ast_{h_r} \cdots
    R^\ast_{h_1} T  \; .
\ee
\vskip.1cm

We obtain indeed a local transformation law since the new
tensor product is designed in such a way that
\be
   (f_1 T_1) \oL (f_2 T_2) = f_1 f_2 \, T_1 \oL T_2
\ee
for all $f_1,f_2 \in \A$ and elements $T_1, T_2$ of tensor
products of $\Omega^1$.

\begin{lemma}
\label{lemma:oLassoc}
The left covariant tensor product $\oL$ is associative:
\be
   (T_1 \oL T_2) \oL T_3 = T_1 \oL (T_2 \oL T_3)
\ee
for all $T_i$ in tensor products of $\Omega^1$.
\end{lemma}
{\bf Proof:} In particular, we find
\bez
     (\theta^{h_1} \oL \theta^{h_2}) \oL T
 &=& (\theta^{h_1} \oA R_{h_1^{-1}}^\ast \theta^{h_2}) \oL T
  = ( \theta^{h_1} \oA \theta^{\ad(h_1^{-1}) h_2} ) \oL T \\
 &=& \theta^{h_1} \oA \theta^{\ad(h_1^{-1}) h_2} \oA
     R^\ast_{[\ad(h_1^{-1}) h_2]^{-1}} \, R^\ast_{h_1^{-1}} T \\
 &=& \theta^{h_1} \oA \theta^{\ad(h_1^{-1}) h_2} \oA
     R^\ast_{h_1^{-1}} \, R^\ast_{h_2^{-1}} T \\
 &=& \theta^{h_1} \oA R^\ast_{h_1^{-1}} \, (\theta^{h_2} \oA
     R^\ast_{h_2^{-1}} T)
  = \theta^{h_1} \oL ( \theta^{h_2} \oL T )  \; .
\eez
Our more general assertion is proved in the same way.
\hfill $\blacksquare$

\begin{lemma}
\label{lemma:Rast_oL}
For all $T_1, T_2$ in tensor products of $\Omega^1$,
\be
   R^\ast_h (T_1 \oL T_2)
 = (R^\ast_h T_1) \oL (R^\ast_h T_2)  \; .
        \label{Rast_oL}
\ee
\end{lemma}
{\bf Proof:}
\bez
   R^\ast_h [ (f \, \theta^{h_1} \oA \ldots \oA \theta^{h_r}) \oL T ]
 &=& R^\ast_h (f \, \theta^{h_1} \oA \ldots \oA \theta^{h_r}) \, R_h^\ast
     R_{h_r^{-1}}^\ast \cdots R_{h_1^{-1}}^\ast
     R_{h^{-1}}^\ast R_h^\ast T \\
 &=& R^\ast_h (f \, \theta^{h_1} \oA \ldots \oA \theta^{h_r}) \,
     R_{[\ad(h) h_r]^{-1}}^\ast \cdots R_{[\ad(h) h_1]^{-1}}^\ast
     R_h^\ast T \\
 &=& R^\ast_h (f \, \theta^{h_1} \oA \ldots \oA \theta^{h_r})
     \oL R^\ast_h T
\eez
for all $f \in \A$ and all $T$ in a tensor product of $\Omega^1$.
Now the assertion follows by linearity.
\hfill $\blacksquare$

\subsection{A left-covariant product in the space of forms}
\label{subsec:cap}
The non-locality of the tensor product $\oA$ discussed above is inherited
by the product in $\Omega$. For a bicovariant group lattice we can define
a left-covariant product in $\Omega$ via
\be
   \omega_1 \cap \omega_2 = \pi ( \omega_1 \oL \omega_2 )
\ee
where $\pi$ is the projection $\Omega \oA \Omega \rightarrow \Omega$.
The new product inherits from $\oL$ left-covariance and
associativity. From the definition we obtain
\be
    (\theta^{h_1} \cdots \theta^{h_r}) \cap \omega
  &=& \theta^{h_1} \cdots \theta^{h_r} \, R^\ast_{h_r^{-1}} \cdots
    R^\ast_{h_1^{-1}} \omega      \label{cap_def}   \\
    \theta^{h_1} \cdots \theta^{h_r} \, \omega
  &=& (\theta^{h_1} \cdots \theta^{h_r}) \cap R^\ast_{h_1} \cdots
    R^\ast_{h_r} \omega
\ee
and also
\be
    (\theta^{h_1} \cap \ldots \cap \theta^{h_r}) \cap \omega
  &=& (\theta^{h_1} \cap \ldots \cap \theta^{h_r}) \, R^\ast_{h_1^{-1}} \cdots
    R^\ast_{h_r^{-1}} \omega  \\
    (\theta^{h_1} \cap \ldots \cap \theta^{h_r}) \, \omega
  &=& (\theta^{h_1} \cap \ldots \cap \theta^{h_r}) \cap R^\ast_{h_r} \cdots
    R^\ast_{h_1} \omega     \; .
\ee
In particular,
\be
    \theta^h \cap \theta^{h'}
  = \theta^h \, \theta^{\ad(h^{-1}) h'} \, , \qquad
    \theta^h \, \theta^{h'}
  = \theta^h \cap \theta^{\ad(h) h'}  \; .
\ee
The 2-form relations (see section 4 of Ref.~\citen{DMH02-gl}) now read
\be
   \sum_{h,h' \in S} \delta^g_{h' h} \, \theta^h
   \cap \theta^{h'} = 0 \qquad  \forall \, g \in S_{(2)}
       \label{2form_rels_cap}
\ee
and a 2-form can be decomposed using the projections
\be
     p_{(e)}(\theta^h \cap \theta^{h'})
 &=& \dl^e_{h h'} \, \theta^h \cap \theta^{h'} \nonumber \\
     p_{(h)}(\theta^{h'}\cap\theta^{h''})
 &=& \dl^h_{h'' h'}\, \theta^{h'} \cap \theta^{h''}
          \qquad \, h \in S_{(1)} \nonumber \\
     p_{(g)}(\theta^h \cap \theta^{h'})
 &=& \dl^g_{h' h} \, \theta^h \cap \theta^{h'}
          \qquad \quad g \in S_{(2)}
\ee
where $S_{(1)} = S^2 \cap S$, $S_{(2)} = S^2 \setminus S_e$ and
$S_e = S \cup \{ e \}$.
 For a cycle $h_1 h_2 = h_2 h_3 = \cdots = h_r h_1$ we obtain
\be
 \theta^{h_1} \theta^{h_2} + \theta^{h_2} \theta^{h_3} + \cdots
 + \theta^{h_r} \theta^{h_1} = \theta^{h_1} \cap \theta^{h_r}
 + \theta^{h_2} \cap \theta^{h_1} + \cdots
 + \theta^{h_r} \cap \theta^{h_{r-1}}   \, .
\ee
Hence the structure of 2-form relations is preserved by the
$\cap$-product.
\vskip.2cm

Since $R_h^\ast$ commutes with $\pi$, (\ref{Rast_oL}) leads to
\be
   R^\ast_h (\omega_1 \cap \omega_2)
 = (R^\ast_h \omega_1) \cap (R^\ast_h \omega_2) \; .
                \label{R^ast_cap}
\ee
\vskip.2cm

In Ref.~\citen{DMH02-gl} a map $\Delta : \Omega \rightarrow \Omega$ has been
introduced which is a graded derivation with respect to the
ordinary product in $\Omega$ and satisfies
\be
    \Delta( \theta^h ) := \sum_{h',h'' \in S} \delta^h_{h'' h'} \,
   \theta^{h'} \cap \theta^{h''} \; .
\ee

\begin{lemma}
\label{lemma:Delta_oL}
$\Delta$ is a graded derivation with respect to the $\cap$-product in $\Omega$.
\end{lemma}
{\bf Proof:} Using (I.4.11) and (I.4.20) we obtain
\bez
     \Delta( \theta^h \cap \omega )
 &=& \Delta( \theta^h \, R^\ast_{h^{-1}} \omega )
  = \Delta(\theta^h) \, R^\ast_{h^{-1}} \omega - \theta^h \, \Delta( R^\ast_{h^{-1}} \omega )
  = \Delta(\theta^h) \, R^\ast_{h^{-1}} \omega - \theta^h \, R^\ast_{h^{-1}} \Delta( \omega ) \\
 &=& ( \sum_{h',h'' \in S} \delta^h_{h'' h'} \, \theta^{h'} \cap \theta^{h''} ) \cap
    R^\ast_{h''h'} \, R^\ast_{h^{-1}} \omega - \theta^h \cap \Delta( \omega )
  = \Delta( \theta^h ) \cap \omega - \theta^h \cap \Delta( \omega )
\eez
for all $\omega \in \Omega$. This in turn implies the general derivation rule
\bez
    \Delta( \omega' \cap \omega )
  = \Delta( \omega' ) \cap \omega + (-1)^r \, \omega' \cap \Delta( \omega )
\eez
where $\omega'$ is an arbitrary $r$-form.
\hfill $\blacksquare$
\vskip.2cm

The map $\d$ is \emph{not} a derivation with respect to
the $\cap$-product. For an $r$-form $\omega$ we
obtain from (I.4.12) the formula
\be
  \d \omega = \sum_{h \in S} \theta^h \cap R_h^\ast \omega
  - (-1)^r \, \omega \cap \theta - \Delta(\omega)
         \label{domega_cap}
\ee
where
\be
      \theta = \sum_{h \in S} \theta^h \; .
\ee
This allows to evaluate $\d$ applied to any form in terms of
expressions which only involve the $\cap$-product (instead of
the original product in $\Omega$).
In fact, we could have defined the left-covariant product of forms
(and moreover the left-covariant tensor product) by its basic properties
(without reference to the tensor product over $\A$) and the action
of $\d$ directly in terms of (\ref{domega_cap}).
Reversing some of the arguments would then demonstrate that there is
a product in $\Omega$ with respect to which $\d$ becomes a derivation.

\subsection{Fixing the ambiguity of 2-form components}
\label{subsec:2form-ambig}
Given a 2-form
\be
   \psi = \sum_{h,h' \in S} \psi_{h,h'} \, \theta^h \cap \theta^{h'}
\ee
the biangle and triangle coefficient functions $\psi_{h,h'}$ are uniquely
determined, but there is an ambiguity in the quadrangle coefficients
as a consequence of the 2-form relations (\ref{2form_rels_cap}). Indeed,
writing
\be
  \psi_{(g)} = p_{(g)} \psi
  = \sum_{h,h' \in S} \check{\psi}_{(g) \, h,h'} \, \theta^h \cap \theta^{h'}
\ee
for $g \in S_{(2)}$, there is a freedom of gauge transformations
$\check{\psi}_{(g) \, h,h'} \mapsto \check{\psi}_{(g) \, h,h'}
+ \Psi_{(g)} \, \delta^g_{h'h}$ with an arbitrary function
$\Psi_{(g)}$ on $G$. \cite{rem:ambig}
For any two members $h,h'$ and $\hat{h}, \hat{h}'$ of the chain
$h_1 {h_1}' = \ldots = h_r {h_r}' = g \in S_{(2)}$, the difference
\be
     \psi_{(g) \, h, h'; \hat{h}, \hat{h}'}
  := \check{\psi}_{(g) \, h,h'} - \check{\psi}_{(g) \, \hat{h},\hat{h}'}
        \label{psidiffs}
\ee
and thus also
\be
    \psi_{(g) \, h',h}
 := \sum_{\hat{h},\hat{h}' \in S} \psi_{(g) \, h', h; \hat{h}', \hat{h}}
  = |g| \, \check{\psi}_{(g) \, h',h}
    - \sum_{\hat{h},\hat{h}' \in S} \dl^g_{\hat{h} \hat{h}'} \,
      \check{\psi}_{(g) \, \hat{h}',\hat{h}}
      \label{psicomp}
\ee
is gauge invariant and hence independent of the
choice of the coefficient functions $\check{\psi}_{(g) \, h,h'}$
(from their gauge equivalence class).
Here $|g|$ denotes the length of the chain which belongs to $g$,
i.e. $|g| = r$. Furthermore, we obtain
$\sum_{h,h'} \dl^g_{h'h} \, \psi_{(g) \, h,h'} = 0$ and
\be
  \psi_{(g)} = {1 \over |g|} \, \sum_{h,h' \in S} \dl^g_{h'h}
  \, \psi_{(g) \, h,h'} \, \theta^h \cap \theta^{h'}
\ee
which suggests to {\em define} the functions (\ref{psicomp})
as the {\em quadrangle components} of the 2-form $\psi$ (with
respect to the $\cap$-product).
The equation $\psi_{(g)} = 0$ (for a 2-form $\psi$) is equivalent
to the vanishing of all the differences
$\psi_{(g) \, h, h'; \hat{h}, \hat{h}'}$.
\vskip.1cm

Of course, also in the case of higher than 2-forms there is an
ambiguity in the choice of coefficients and a corresponding
way of fixing it.

\subsection{Left-covariant metric and compatibility with a linear connection}
\label{subsec:compat}
Let us express $\mathsf{g}$ given in (\ref{g_oA}) as
\be
   \mathsf{g} = \sum_{h,h' \in S} \g_{h,h'} \, \theta^h \oL \theta^{h'}
                \label{g_oL}
\ee
with $\g_{h,h'} \in \A$. By comparison with (\ref{g_oA}), we obtain
\be
    \gamma_{h,h'} = \g_{h,hh'h^{-1}}  \; .
\ee
We say that $\mathsf{g}$ is {\em symmetric} if $\g_{h,h'} = \g_{h',h}$, which
corresponds to $\gamma_{h,h^{-1}h'h} = \gamma_{h',h'{}^{-1}hh'}$.
Furthermore, $\mathsf{g}$ is said to be {\em invertible} if the matrix
$\g = (\g_{h,h'})$ is invertible (at all sites).
\vskip.1cm

An object $\mathsf{g}$ as considered above is a candidate for a ``metric tensor''.
Its components should then be expected to determine lengths of vectors
and angles between vectors at a site. This interpretation clearly
distinguishes the components $\g_{h,h'}$ and thus the left-covariant
tensor product (see also the corresponding remarks in the introduction).
Hence we define a {\em metric tensor} as an object $\mathsf{g}$
of the form (\ref{g_oL}) such that the coefficient matrix $\g$
is real, symmetric and invertible.
\vskip.1cm

A metric $\mathsf{g}$ is called \emph{left-invariant} if
$L_h^\ast \mathsf{g} = \mathsf{g}$ for all $h \in S$,
where $L_h$ denotes the left action by $h$ on $G$. This is
equivalent to a ``constant metric'', i.e. $\g_{h,h'} \in \mathbb{R}$.
A left-invariant metric is called \emph{bi-invariant} if it is also
\emph{right-invariant}, i.e. $R_h^\ast \mathsf{g} = \mathsf{g}$
for all $h \in S$. This means that the metric is constant and satisfies
$\g_{h_1,h_2} = \g_{\ad(h)h_1 , \ad(h)h_2}$ for all $h,h_1,h_2 \in S$.
\vskip.1cm

Let $\{ \ell_h \, | \, h \in S \}$ be the vector fields dual to
$\{ \theta^h \, | \, h \in S \}$, so that $\ell_h f = R_h^\ast f - f$
for $f \in \A$. Let $\V_{\ell_h}$ be the parallel
transport along the vector field $\ell_h$ with respect to a
linear connection on $\Omega^1$ (see Ref.~\citen{DMH02-gl}). We write
\be
   \V_{\ell_{h'}} \theta^h
 = \sum_{h'' \in S} (R^\ast_{{h'}^{-1}} V^h{}_{h',h''}) \, \theta^{h''}
        \label{Vellh_thetah}
\ee
where $V_h = (V^{h''}{}_{h,h'})$ are matrices with entries in $\A$.
$\V$ extends to $\Omega^1 \oL \Omega^1$ via
\be
   \V_{\ell_h}(\alpha \oL \beta )
 = \V_{\ell_h} \alpha \oL \V_{\ell_h} \beta  \; .
       \label{Vellh_oL}
\ee
Then $\V := \sum_{h \in S} \theta^h \oA \V_{\ell_h}$ has the
property $\V(f \, \alpha \oL \beta) = f \, \V(\alpha \oL \beta)$
and thus defines a connection according to lemma~6.1 of
Ref.~\citen{DMH02-gl}.
\vskip.2cm

An element $\mathsf{g} \in \Omega^1 \oL \Omega^1$ (e.g., a metric) is said to
be {\em compatible} with the linear connection $\na$ if
\be
     \na \mathsf{g} = 0
\ee
which in terms of the parallel transport operators takes the form
\be
   \V_{\ell_h} \mathsf{g} = \mathsf{g}  \qquad \forall h \in S \; .
\ee
In components, this reads
\be
   \sum_{h_1,h_2 \in S} V^{h_1}{}_{h,h'_1} \, V^{h_2}{}_{h,h'_2}
   \, \g_{h_1,h_2} = R^\ast_h \, \g_{h'_1,h'_2}
\ee
and in matrix form
\be
    R^\ast_h \g = V_h^T \, \g \, V_h  \; .
             \label{Vg_compat}
\ee
If $\mathsf{g}$ is a metric, this condition requires that the matrices
$V_h$, $h \in S$, are invertible. For a given metric, there
are not always matrices $V_h$ satisfying (\ref{Vg_compat}).

\begin{lemma}
A linear connection compatible with a metric on a bicovariant group lattice
exists if and only if the metric has the same signature at all sites.
\end{lemma}
{\bf Proof:} This is a direct consequence of the fact that two real
symmetric matrices $A,B$ with the same rank are related by $B = V^T A V$
with an invertible matrix $V$ if and only if both have the same signature.
\hfill $\blacksquare$

\vskip.2cm
A bicovariant group lattice supplied with a metric of constant signature
will be called a \emph{Riemannian group lattice} in the following.
Since we require a metric to be non-degenerate, a Riemannian group
lattice $(G,S,\mathsf{g})$ should be regarded as an $|S|$-dimensional
structure.
\vskip.1cm

The metric-compatibility condition determines the transport matrices,
and thus the connection, only up to transformations $V_h \mapsto J_h \, V_h$
with arbitrary isometries $J_h$, which are matrices of functions on $G$
such that
\be
    J_h^T \, \g \, J_h = \g  \; .  \label{J-isom}
\ee

\subsection{Backward parallel transport of vector fields and
 geometric interpretation of metric-compatible linear
 connections}
\label{subsec:mc-conn-interpr}
Vector fields are elements of the $\A$-bimodule generated by
$\{ \ell_h \, | \, h \in S \}$ \cite{DMH02-gl}.
A linear connection determines a backward parallel transport of
vector fields along a vector field:
\be
   \tilde{\V}_{\ell_h} X
 := \sum_{h',h'' \in S} (R_h^\ast X^{h'}) \, V^{h''}{}_{h,h'}
    \cdot \ell_{h''} \, , \qquad
    \tilde{\V}_X := \sum_{h \in S} X^h \, \tilde{\V}_{\ell_h}
        \label{tVell-ell}
\ee
(see Ref.~\citen{DMH02-gl} for details). The vectors
\be
   V_{h,h'} := \tilde{\V}_{\ell_h} \ell_{h'}
                = \sum_{h'' \in S} V^{h''}{}_{h,h'} \cdot \ell_{h''}
\ee
are the images in the tangent space at $g$ of the vectors
$\ell_{h'}$ at $gh$. If the transport is metric-compatible,
the vectors $V_{h,h'}$ at $g$ carry the metric properties
of $\ell_{h'}$ at $gh$, i.e.
\be
   \g_{h',h''}(g h) = \mathsf{g}(\ell_{h'},\ell_{h''})(g h)
  = \mathsf{g}(V_{h,h'},V_{h,h''})(g) \; .
\ee
\vskip.1cm

Of course, we can also transport tangent vectors from more remote sites
to the tangent space at $g$ by iterated application of the operators
$\tilde{\V}_{\ell_h}$:
\be
   V_{h_1,\ldots,h_{r+1}} := \tilde{\V}_{\ell_{h_1}} \cdots
       \tilde{\V}_{\ell_{h_r}} \ell_{h_{r+1}} \; .
\ee
The results will, however, be path-dependent
in general. But here we see very clearly the geometric significance
of a metric-compatible linear connection. It maps part of the group
lattice into the tangent space at a site in such a way that the
metric relations are preserved, i.e. isometrically. In general, this
cannot be done for the whole group lattice.
Torsion and curvature are obstructions.
We have already shown in Ref.~\citen{DMH02-gl} that in case of
vanishing torsion at least the next-neighbor part of the group lattice
is mapped isometrically into the tangent space in this way, i.e.
the backward parallel transport preserves the group lattice geometry
to first order. Curvature is a second order obstruction.
Its biangle, triangle and quadrangle parts are given, respectively, by
\be
\begin{array}{r@{\, = \,}l@{\quad\mbox{if}\quad}l}
 ( \tilde{\V}_{\ell_{h_1}} \tilde{\V}_{\ell_{h_2}} - I ) \, \ell_{h}
    & ( V_{h_1} \, R^\ast_{h_1} V_{h_2} - I )^{h'}{}_h \, \ell_{h'}
    &  h_1h_2 = e                \\
 ( \tilde{\V}_{\ell_{h_1}} \tilde{\V}_{\ell_{h_2}} - \tilde{\V}_{\ell_{h_3}} ) \, \ell_{h}
    & ( V_{h_1} \, R^\ast_{h_1} V_{h_2} - V_{h_3} )^{h'}{}_h \, \ell_{h'}
    &  h_1 h_2 = h_3 \in S_{(1)}  \\
 ( \tilde{\V}_{\ell_{h_1}} \tilde{\V}_{\ell_{h_2}}
   - \tilde{\V}_{\ell_{\hat{h}_1}} \tilde{\V}_{\ell_{\hat{h}_2}} ) \, \ell_{h}
    & ( V_{h_1} \, R^\ast_{h_1} V_{h_2}
      - V_{\hat{h}_1} \, R^\ast_{\hat{h}_1} V_{\hat{h}_2} )^{h'}{}_h \, \ell_{h'}
    & h_1 h_2 = \hat{h}_1 \hat{h}_2 \in S_{(2)} \, .
    \end{array}  \label{curv&ptransport}
\ee
An equivalent curvature definition will be presented in
section~\ref{sec:tor&curv}.
The last formula of (\ref{curv&ptransport}) is a discrete version of a
familiar formula of continuum differential geometry: the quadrangle
curvature is determined by parallel transport of a vector field
around a quadrangle. There are no counterparts of biangle and
triangle curvature in continuum differential geometry.
\vskip.1cm

Let us make more precise how an isometric tangent space picture
of (part of) a group lattice is obtained if a metric-compatible
linear connection is given.
If $S$ has $n$ different elements, let $(\, , \, )$ be an inner product
in $\mathbb{R}^n$ with the same signature as $\mathsf{g}$. At the origin in
$\mathbb{R}^n$ we choose an $n$-bein $\{ \mathbf{u}_h \, | \, h \in S \}$
such that
\be
    (\mathbf{u}_h, \mathbf{u}_{h'}) = \mathsf{g}(\ell_h,\ell_{h'})(g)  \; .
\ee
Then $\iota : \ell_h \mapsto \mathbf{u}_h$ extends to an isomorphism of metric
linear spaces. Furthermore, $\mathbf{V}_{h,h'} := \iota(V_{h,h'})
  = \sum_{h'' \in S} V^{h''}{}_{h,h'} \, \mathbf{u}_{h''}$ represents the
vector $V_{h,h'}$ in $\mathbb{R}^n$. We attach it at the tip of
$\mathbf{u}_h$. More generally, the vector
\be
    \mathbf{V}_{h_1,\ldots,h_{r+1}} := \iota( V_{h_1,\ldots,h_{r+1}} )
  = \sum_{h \in S} \mathbf{u}_h \, [V_{h_1}(g) \, V_{h_2}(g h_1) \cdots
    V_{h_r}(gh_1\cdots h_{r-1})]^h{}_{h_{r+1}}
      \label{V_h_general}
\ee
has to be attached at the tip of
$\mathbf{u}_{h_1} + \mathbf{V}_{h_1,h_2} + \ldots + \mathbf{V}_{h_1,\ldots,h_r}$.
\vskip.1cm

The isometries $J_h$ act on the vectors $\mathbf{V}_{h,h'}$ as follows,
\be
    J_h( \mathbf{V}_{h,h'} ) := \sum_{h_1,h_2} \mathbf{u}_{h_1} \,
       J^{h_1}{}_{h,h_2} \, V^{h_2}{}_{h,h'} \; .
                 \label{JhV-vector}
\ee
The isometry property of the $J_h$ then implies
\be
    ( J_h( \mathbf{V}_{h,h'} ) , J_h( \mathbf{V}_{h,h''} ) )
  = ( \mathbf{V}_{h,h'} , \mathbf{V}_{h,h''} )  \; .
                 \label{JhV-isom}
\ee
The backward parallel transport and the isomorphism $\iota$ provide
us with a convenient way to describe the action of a (metric-compatible)
linear connection in $\mathbb{R}^n$ (supplied with a standard inner
product). This will be used extensively in sections \ref{sec:discrRiem}
and \ref{sec:gl-torsion}.

\subsection{Contravariant metric tensor and compatibility with a
 linear connection}
A left-covariant tensor product of vector fields $X,Y$ is
defined as follows,
\be
  X \oL Y := \sum_{h \in S} X^h \, \ell_h \oA R_{h \ast} Y \, .
\ee
Given a metric tensor in the sense of section~\ref{subsec:compat},
there is also a ``contravariant'' metric tensor,
\be
 \mathsf{h} = \sum_{h,h' \in S} \h^{h,h'} \cdot \ell_h \oL \ell_{h'}
    = \sum_{h,h' \in S} \h^{h,\ad(h)h'} \cdot \ell_h \oA \ell_{h'}
\ee
where $(\h(g)^{h,h'})$ is the inverse of the matrix $\g$
at $g \in G$.
\vskip.1cm

If the matrices $V_h$ are invertible for all $h \in S$, the
corresponding linear connection on $\Omega^1$ induces a connection
on the space $\X$ of vector fields (see Ref.~\citen{DMH02-gl}).
An element $\mathsf{h} \in \X \oL \X$ is {\em compatible} with the
connection $\na$ if
\be
     \na \mathsf{h} = 0
\ee
where $\na$ has been extended to $\X \oL \X$ following
the procedure in section~\ref{subsec:compat}.
Using $U_h := V_h^{-1}$, this condition reads
\be
  R^\ast_h \h^{h_1,h_2} =
  \sum_{h'_1,h'_2 \in S} (U_h)^{h_1}{}_{h'_1} \, (U_h)^{h_2}{}_{h'_2}
  \, \h^{h'_1,h'_2}
\ee
or $R^\ast_h \h = U_h \, \h \, U_h^T$ in matrix form.

\subsection{Discrete Killing vector fields}
Let $X = \sum_{h \in S} X^h \cdot \ell_h$ be a discrete vector field
for which the map $\phi_X : G \rightarrow G$, which is determined by
$\phi_X^\ast = I+X$ on functions, is differentiable
(see Ref.~\citen{DMH02-gl}).
$X$ will be called a {\em Killing vector field} of a metric $\mathsf{g}$ if
$\mbox{\pounds}_X \mathsf{g} = \phi^\ast_X \mathsf{g} - \mathsf{g} =0$ (with the Lie
derivative \pounds{} introduced in Ref.~\citen{DMH02-gl}).
For $X = \ell_h$ this becomes $R_h^\ast \mathsf{g} = \mathsf{g}$, i.e.
\be
  \g(g h)_{h_1,h_2} = \g(g)_{\ad(h) h_1 , \ad(h) h_2}
                         \label{Killing}
\ee
for all $g \in G$. The right hand side of (\ref{Killing}) can
be expressed in the form $( P_h^T \, \g(g) \, P_h )_{h_1,h_2}$
where the matrix $P_h$ represents a permutation.
\vskip.1cm

A metric $\mathsf{g}$ on a bicovariant group lattice $(G,S)$ is thus
right-invariant if it satisfies $\mbox{\pounds}_{\ell_h} \mathsf{g} = 0$
for all $h \in S$.
A right-invariant metric is completely determined by its
values at one site (e.g., at the unit element $e$).
\vskip.1cm

$V_h := P_h$ defines a linear connection which is compatible with every
right-invariant metric. Each other linear connection compatible with a
right-invariant metric is then obtained as $V_h := J_h P_h$, where $J_h$ is
at each lattice site an isometry of the metric.
\vskip.2cm

\begin{example}
\label{ex:Zn-Killing}
Let $G$ be a discrete group and $S \subset G \setminus \{ e \}$ finite
and \emph{Abelian}. If $\mbox{\pounds}_{\ell_h} \mathsf{g} = 0$
for some $h \in S$, the condition (\ref{Killing}) becomes
$\g(g h) = \g(g)$ which means that the functions
$\g_{h_1,h_2}$, $h_1,h_2 \in S$, are constant on the orbits in $G$
under the right action $R_h$.

Let $G = \mathbb{Z}_n$ or $G = \mathbb{Z}$, and $1 \in S$. If $\ell_1$
is a Killing vector field, the metric coefficients $\g_{h_1,h_2}$ are
constant on the whole group. The corresponding natural linear connection
is then given by $V_h = I$.
\end{example}

\begin{example}
\label{ex:S3-Killing}
Let $G = {\cal S}_3$ and $S = \{ (12),(13),(23) \}$.
If $\ell_{(12)}$ is a Killing vector field of a metric $\mathsf{g}$ on
this group lattice, then
\be
   \g(g \, (12)) = P_{(12)} \, \g(g) \, P_{(12)}
 \quad \mbox{where} \quad
  P_{(12)} = \left( \begin{array}{ccc}
               1 & 0 & 0 \\
               0 & 0 & 1 \\
               0 & 1 & 0 \end{array} \right) \, .
\ee
This determines the metric at the sites $(12),\,(13),\,(23)$
in terms of the metric at the sites $e,\,(132),\,(123)$, respectively.
If $\ell_{(13)}$ and $\ell_{(23)}$ are Killing vector fields of
$\mathsf{g}$, then
\be
     \g(g \, (13)) &=& P_{(13)} \, \g(g) \, P_{(13)}
 \quad \mbox{where} \quad
  P_{(13)} = \left( \begin{array}{ccc}
               0 & 0 & 1 \\
               0 & 1 & 0 \\
               1 & 0 & 0 \end{array} \right) \nonumber \\
     \g(g \, (23)) &=& P_{(23)} \, \g(g) \, P_{(23)}
 \quad \mbox{where} \quad
  P_{(23)} = \left( \begin{array}{ccc}
               0 & 1 & 0 \\
               1 & 0 & 0 \\
               0 & 0 & 1 \end{array} \right) \; .
\ee
The right-invariant metrics on $({\cal S}_3, \{ (12),(13),(23) \})$
are then given by
\be
   \g(e) &=& \left( \begin{array}{ccc}
               a & b & c \\
               b & d & e \\
               c & e & f \end{array} \right) \, , \;
   \g((12)) = \left( \begin{array}{ccc}
               a & c & b \\
               c & f & e \\
               b & e & d \end{array} \right) \, , \;
   \g((13)) = \left( \begin{array}{ccc}
               f & e & c \\
               e & d & b \\
               c & b & a \end{array} \right)  \nonumber \\
   \g((23)) &=& \left( \begin{array}{ccc}
               d & b & e \\
               b & a & c \\
               e & c & f \end{array} \right) \, , \;
   \g((123)) = \left( \begin{array}{ccc}
               d & e & b \\
               e & f & c \\
               b & c & a \end{array} \right) \, , \;
   \g((132)) = \left( \begin{array}{ccc}
               f & c & e \\
               c & a & b \\
               e & b & d \end{array} \right)  \quad
\ee
with constants $a,b,c,d,e,f$. A linear connection compatible with
this family of metrics is obtained by choosing $V_h = P_h$.
The family of right-invariant metrics includes the following
bi-invariant metric:
\be
   \g(h) = \left( \begin{array}{ccc}
               a & b & b \\
               b & a & b \\
               b & b & a \end{array} \right)
\ee
with constants $a,b$.
\end{example}

\section{Torsion and curvature of linear connections as left-covariant
         tensors on bicovariant group lattices}
\label{sec:tor&curv}
\setcounter{equation}{0}
The torsion 2-forms
\be
  \Theta^h
  = \d \theta^h - \pi (\nabla \theta^h)
  = \d \theta^h - \theta \, \theta^h
  - \sum_{h',h'' \in S} V^h{}_{h',h''} \, \theta^{h'} \theta^{h''}
\ee
can be rewritten in terms of the $\cap$-product and then decomposed
into biangle, triangle and quadrangle parts as follows,
\be
     \Theta^h
 &=& \sum_{h_1,h_2 \in S} Q^h{}_{h_1,h_2} \, \theta^{h_1} \cap \theta^{h_2}
          \nonumber \\
 &=& \sum_{h_1,h_2 \in S} \Big( Q^h_{(e) \, h_1,h_2}
     + \sum_{h_0 \in S_{(1)}} Q^h_{(h_0) \, h_1,h_2}
     + \sum_{g \in S_{(2)}} \check{Q}^h_{(g) \, h_1,h_2} \Big)
     \, \theta^{h_1} \cap \theta^{h_2} \; . \label{Theta_Q}
\ee
In this way we find the biangle components
\be
  Q^h_{(e) \, h_1,h_2} = \dl^e_{h_1h_2} (\dl^h_{h_1} + V^h{}_{h_1,h_2})
     \label{biangle_torsion}
\ee
and the triangle components
\be
   Q^h_{(h_0) \, h_1,h_2}
 = \dl^{h_0}_{h_2 h_1} (\dl^h_{h_1} - \dl^h_{h_0}
   + V^h{}_{h_1,h_1^{-1} h_2 h_1}) \; .
     \label{triangle_torsion}
\ee
In case of the quadrangle components, one has to take the 2-form
relations (\ref{2form_rels_cap}) into account. As a consequence of
the latter, the functions $Q^h_{(g) \, h_1,h_2}$ are not uniquely
determined. Following the discussion in section~\ref{subsec:2form-ambig},
it is convenient to introduce the differences
\be
      Q^h_{(g) \, h_1,h_2;\hat{h}_1,\hat{h}_2}
 &:=& \check{Q}^h_{(g) \, h_1,h_2} - \check{Q}^h_{(g) \,
      \hat{h}_1,\hat{h}_2} \nonumber \\
  &=& \dl^g_{h_2h_1} (\dl^h_{h_1} - \dl^h_{\hat{h}_1}
      + V^h{}_{h_1,h_1^{-1} h_2 h_1}
      - V^h{}_{\hat{h}_1,\hat{h}_1^{-1} \hat{h}_2 \hat{h}_1})
      \label{quad_tor_diffs}
\ee
(cf. (\ref{psidiffs}))
where $\hat{h}_2, \hat{h}_1$ is any pair of elements of $S$
which belongs to the same chain as $h_2,h_1$ (so that
$\hat{h}_2 \hat{h}_1 = g = h_2 h_1$). In particular, the vanishing
of the quadrangle part of the torsion 2-form is equivalent to
the vanishing of all the quantities (\ref{quad_tor_diffs}).
According to section~\ref{subsec:2form-ambig}, the quadrangle torsion
components should be defined as follows,
\be
       Q^h_{(g) \, {h_i}',h_i}
  := |g| \, \check{Q}^h_{(g) \, {h_i}',h_i}
      - \sum_{h',h'' \in S} \dl^g_{h'h''} \, \check{Q}^h_{(g) \, h'',h'}
   = \sum_{h',h'' \in S \atop h'h''=g} Q^h_{(g) \, {h_i}', h_i; h'', h'}
      \quad  i=1, \ldots, |g|    \label{quad_tor_comp}
\ee
if $h_1 {h_1}' = \ldots = h_r {h_r}' = g$ is the corresponding chain.
This does not depend on the choice of the coefficient functions
$\check{Q}^h_{(g) \, h',h}$ which is ambiguous as a consequence
of the 2-form relations.
\vskip.2cm

After some manipulations like
\be
 & & \sum_{h',h'' \in S} \theta^{h'} \theta^{h''} \oA \V_{\ell_{h''}}
     \V_{\ell_{h'}} \theta^h   \nonumber \\
 &=& \sum_{h_1,h'' \in S} (\theta^{h_1} \theta^{h''}) \oL R_{h_1}^\ast
     R_{h''}^\ast \V_{\ell_{h''}} \V_{\ell_{h_1}} \theta^h
           \nonumber \\
 &=& \sum_{h_1,h'' \in S} (\theta^{h_1} \cap \theta^{\ad(h_1)h''}) \oL R_{h_1}^\ast
     R_{h''}^\ast \V_{\ell_{h''}} \V_{\ell_{h_1}} \theta^h
           \nonumber \\
 &=& \sum_{h_1,h_2 \in S} (\theta^{h_1} \cap \theta^{h_2}) \oL R_{h_2 h_1}^\ast
     \V_{\ell_{\ad(h_1^{-1}) h_2}} \V_{\ell_{h_1}} \theta^h
           \nonumber \\
 &=& \sum_{h_1,h_2 \in S} (\theta^{h_1} \cap \theta^{h_2}) \oL R_{h_2 h_1}^\ast
     \sum_{h',h''} (R_{(h_2 h_1)^{-1}}^\ast V^h_{h_1,h''}) \,
     (R_{h_1^{-1} h_2^{-1} h_1}^\ast V^{h''}_{h_1^{-1} h_2 h_1,h'}) \, \theta^{h''}
           \nonumber \\
 &=& \sum_{h_1,h_2,h',h'' \in S} (\theta^{h_1} \cap \theta^{h_2}) \oL
     V^h_{h_1,h''} \, (R_{h_1}^\ast V^{h''}_{h_1^{-1} h_2 h_1,h'}) \,
     \theta^{\ad(h_2h_1) h''}
           \nonumber \\
 &=& \sum_{h_1,h_2,h',h'' \in S} (\theta^{h_1} \cap \theta^{h_2}) \oL
     V^h_{h_1,h''} \,
     (R_{h_1}^\ast V^{h''}_{\ad(h_1^{-1}) h_2,\ad[(h_2 h_1)^{-1}] h'}) \,
     \theta^{h'}
\ee
the definition of the curvature, see (I.7.4), leads to
\be
     \R(\theta^h)
 &=& \sum_{h',h_1,h_2 \in S} \Big( \sum_{h'' \in S} V^h{}_{h_1,h''}
     \, (R^\ast_{h_1} V^{h''}{}_{h_1^{-1} h_2 h_1,\ad[(h_2h_1)^{-1}]h'})
                        \nonumber  \\
 & & - \dl_{h'}^h \, \dl^e_{h_2h_1}
     - \sum_{h'' \in S} \dl^{h''}_{h_2h_1} V^h{}_{h'',\ad[(h_2h_1)^{-1}]h'} \Big)
     \, \theta^{h_1} \cap \theta^{h_2} \oL \theta^{h'}  \; .
\ee
Writing
\be
     \R(\theta^h)
 &=& \sum_{h',h_1,h_2} \R^h{}_{h',h_1,h_2} \,
     \theta^{h_1} \cap \theta^{h_2} \oL \theta^{h'}
                      \nonumber \\
 &=& \sum_{h_1,h_2 \in S} \Big( \R^h_{(e) \,h', h_1, h_2}
     + \sum_{h_0 \in S_{(1)}} \R^h_{(h_0) \, h', h_1, h_2}
     + \sum_{g \in S_{(2)}} \check{\R}^h_{(g) \, h', h_1, h_2} \Big) \,
       \theta^{h_1} \cap \theta^{h_2} \oL \theta^{h'}  \qquad
\ee
we obtain the biangle components
\be
   \R^h_{(e) \, h',h_1,h_2}
 = \dl^e_{h_2h_1} \, ( V_{h_1} \, R^\ast_{h_1} V_{h_2} - I )^h{}_{h'}
   \; ,          \label{biangle_curv}
\ee
the triangle components
\be
   \R^h_{(h_0) \, h',h_1,h_2}
 = \dl^{h_0}_{h_2h_1} \, ( V_{h_1} \, R^\ast_{h_1} V_{h_1^{-1}h_2h_1}
   - V_{h_0} )^h{}_{h_0^{-1}h'h_0} \, ,    \label{triangle_curv}
\ee
and the differences of quadrangle components
\be
      \R^h_{(g) \, h',h_1,h_2; \hat{h}_1, \hat{h}_2}
 &:=& \check{\R}^h_{(g) \, h',h_1,h_2 }
      - \check{\R}^h_{(g) \, h', \hat{h}_1, \hat{h}_2} \nonumber \\
 &=& \dl^g_{h_2h_1} \, ( V_{h_1} \, R^\ast_{h_1} V_{h_1^{-1}h_2h_1}
     - V_{\hat{h}_1} \, R^\ast_{\hat{h}_1}
     V_{\hat{h}_1^{-1} \hat{h}_2 \hat{h}_1} )^h{}_{g^{-1} h'g}
     \; .       \label{quadrangle_curv}
\ee
Again, $\hat{h}_2, \hat{h}_1$ is any pair with
$\hat{h}_2 \hat{h}_1 = g \in S_{(2)}$.
\vskip.2cm

According to section~\ref{subsec:2form-ambig}, the quadrangle curvature
components should be defined as follows,
\be
      \R^h_{(g) \, h', {h_i}', h_i}
 &:=& |g| \, \check{\R}^h_{(g) \, h', h_i', h_i}
      - \sum_{h'',h''' \in S} \dl^g_{h''h'''} \,
        \check{\R}^h_{(g) \, h', h''',h''} \nonumber \\
 &=& \sum_{ h'',h''' \in S \atop h'' h''' = g }
       \R^h_{(g) \, h', {h_i}', h_i; h''', h''}
       \qquad  i=1, \ldots, |g|   \label{quad_curv_comp}
\ee
if $h_1 {h_1}' = \ldots = h_r {h_r}' = g$ is the corresponding chain.
Understanding that the quadrangle part of $\R^h{}_{h',h'',h'''}$ is
given by the above expression, the components of a {\em Ricci tensor}
can be defined without ambiguity as follows,
\be
   {\it Ric}_{h,h'} := \sum_{h'' \in S} \R^{h''}{}_{h,h'',h'} \; .
     \label{Riccit1}
\ee
With the help of a metric, a {\em curvature scalar} can be built:
\be
 R := \sum_{h, h' \in S} (\g^{-1})^{ h, h'} {\it Ric}(\ell_h,\ell_{h'})
\ee
There is, however, another contraction of the curvature tensor, namely
\be
   \widetilde{\it Ric}_{h,h'} := \sum_{h'' \in S} \R^{h''}{}_{h,h',h''}
\ee
which leads in general to a different Ricci tensor and curvature
scalar. Moreover, also the contraction $\sum_{h'' \in S} \R^{h''}{}_{h'',h,h'}$
is in general different from zero. This complicates finding a suitable
analogue of the Einstein equation, for example.

\subsection{Bianchi identities}
According to Ref.~\citen{DMH02-gl}, the first Bianchi identity can be expressed
as follows,
\be
    \d \Theta^h + \Theta( \nabla \theta^h ) = \pi \circ \R( \theta^h )
  = \sum_{h',h_1,h_2 \in S} \R^h{}_{h',h_1,h_2} \,
     \theta^{h_1} \cap \theta^{h_2} \cap \theta^{h'} \;.  \label{1stBianchi}
\ee
Using $\theta \, \omega = \sum_{h \in S} \theta^h \cap R_h^\ast \omega$ we find
\be
   \Theta( \nabla \theta^h )
 &=& - \theta \, \Theta^h + \sum_{h',h'' \in S} V^h{}_{h',h''} \, \theta^{h''} \, \Theta^{h'}
      \nonumber \\
 &=& - \sum_{h' \in S} \theta^{h'} \cap R_{h'}^\ast \Theta^h
   + \sum_{h',h'' \in S} V^h{}_{h',h''} \, \theta^{h''} \cap R_{h''}^\ast \Theta^{h'}
\ee
and thus, with the help of (\ref{domega_cap}),
\be
    \d \Theta^h + \Theta( \nabla \theta^h )
  = - \Theta^h\cap \theta - \Delta(\Theta^h)
    + \sum_{h',h''} V^h{}_{h',h''}\theta^{h''}\cap (R^\ast_{h''}\Theta^{h'}) \; .
\ee
Replacing the left hand side of (\ref{1stBianchi}) with the last expression,
we obtain the first Bianchi identity in terms of the $\cap$-product.
In case of vanishing torsion, it reduces to
\be
  \sum_{h',h_1,h_2 \in S} \R^h{}_{h',h_1,h_2}
  \, \theta^{h_1} \cap \theta^{h_2} \cap \theta^{h'} = 0 \; .
\ee
\vskip.1cm

Using $V^h{}_{h'} := \sum_{h'' \in S} V^h{}_{h'',h'} \, \theta^{h''}$ and
\be
   \R^h{}_{h'} := \sum_{h_1,h_2 \in S} \R^h{}_{\ad(h_2h_1)h',h_1,h_2}
                 \, \theta^{h_1} \cap \theta^{h_2}
\ee
the second Bianchi identity (I.7.15) reads
\be
    \Delta ( \R^h{}_{h'} )
 &=& \sum_{h'' \in S} ( V^h{}_{h''} \, \R^{h''}{}_{h'}
     - \R^h{}_{h''} \, V^{h''}{}_{h'} )
  = \sum_{h_1,h_2 \in S} V^h{}_{h_1,h_2} \,
     \theta^{h_1} \cap R^\ast_{h_1} \R^{h_2}{}_{h'}  \nonumber \\
 & & - \sum_{h_1,h_2,h_3 \in S} \R^h{}_{\ad(h_3h_2)h_1,h_2,h_3}
       \, \theta^{h_2} \cap \theta^{h_3} \cap R^\ast_{h_3 h_2}
       V^{h_1}{}_{h'}  \; .
\ee
Evaluating the left hand side with the help of lemma~\ref{lemma:Delta_oL},
this yields a three-form expression which only involves the $\cap$-product.

\subsection{Integrability conditions of the metric-compatibility equation}
The integrability condition for the metric-compatibility of a linear
connection is $\na^2 \mathsf{g} = 0$ and thus involves the curvature.
After some manipulations we obtain the conditions
\be
 & V_{h_1} R^\ast_{h_1} V_{h_2} = B_{h_1,h_2}  \quad
   & \mbox{for a biangle $h_1h_2=e$}                \\
 & V_{h_1} R^\ast_{h_1} V_{h_2} = T_{h_1,h_2} \, V_h
   & \mbox{for a triangle $h_1h_2=h \in S_{(1)}$}   \\
 & V_{h_1} R^\ast_{h_1} V_{h_2} = K_{h_1,h_2; \hat{h}_1 \hat{h}_2}
     \, V_{\hat{h}_1} R^\ast_{\hat{h}_1} V_{\hat{h}_2}
   & \mbox{for a quadrangle $h_1h_2 = \hat{h}_1 \hat{h}_2 \in S_{(2)}$}
\ee
where for all $g \in G$ the matrices $B_{h_1,h_2}(g), \, T_{h_1,h_2}(g), \,
K_{h_1,h_2; \hat{h}_1 \hat{h}_2}(g)$ are elements of the
isometry group of $\g(g)$. Now we obtain for biangles
\be
 \R^h{}_{(e) \, h',h_1,h_2} = \dl^e_{h_2h_1} \, ( B_{h_1,h_2} - I )^h{}_{h'} \, ,
\ee
for triangles
\be
 \R^h{}_{(h_0) \, h',h_1,h_2} = \dl^{h_0}_{h_2h_1} \, \Big( (T_{h_1,h_2}-I)
 V_{h_0} \Big)^h{}_{h_0^{-1}h'h_0} \, ,  \label{integr_mc_triangle}
\ee
and for quadrangles
\be
   \R^h{}_{(g) \, h',h_1,h_2;\hat{h}_1,\hat{h}_2}
 = \dl^g_{h_2h_1} \Big( (K_{h_1,h_2;\hat{h}_1,\hat{h}_2}-I)
   \, V_{\hat{h}_1} R^\ast_{\hat{h}_1} V_{\hat{h}_1^{-1} \hat{h}_2 \hat{h}_1}
   \Big)^h{}_{g^{-1}h'g}
\ee
where $\hat{h}_2 \hat{h}_1 = g \in S_{(2)}$. As a consequence,
the essential part of the curvature tensor is given by the isometries $B,T,K$.

\subsection{Torsion and curvature as maps on vector fields}
\label{subsec:curv&tor-vf}
Let $(G,S)$ be a bicovariant group lattice and $Q^h{}_{h_1,h_2}$
the torsion tensor components introduced in (\ref{Theta_Q}) with
the quadrangle part defined in (\ref{quad_tor_comp}). For
vector fields $X,Y$ we introduce the {\em torsion tensor}
\be
  Q(X,Y) := \sum_{h \in S, \, h_1,h_2 \in S_e} X^{h_1} Y^{h_2} \,
            Q^h{}_{h_1,h_2} \cdot \ell_h     \; .
\ee
This expression obviously satisfies
$Q(f \cdot X, f' \cdot Y) = f f' \, Q(X,Y)$
and is therefore a (left) tensor.
In the following we consider in more detail the case where $X,Y$
are basic. The torsion tensor can then be written as
\be
  Q(X,Y) = \sum_{h \in S} Q^h{}_{s_X,s_Y} \cdot \ell_h
\ee
where the map $s_X \, : \, G \rightarrow S$ is determined by
$X^h(g) = \delta^h_{s_X(g)}$ (see also Ref.~\citen{DMH02-gl}).
\vskip.2cm

Below we will need the following expression for basic vector fields $X,Y$,
\be
     \tilde{\V}_X R_{X \ast} Y
 &=& \sum_{h_1,h_2 \in S} X^{h_1} \, Y^{\ad(h_1)h_2}
     \, \tilde{\V}_{\ell_{h_1}} \ell_{h_2}
  = \sum_{h_1,h_2 \in S} \delta^{h_1,s_X} \,
     \delta^{h_2,\ad(s_X)^{-1}s_Y}
     \, \tilde{\V}_{\ell_{h_1}} \ell_{h_2}  \nonumber \\
 &=& \sum_{h \in S} V^h{}_{s_X,\ad(s_X)^{-1}s_Y} \cdot \ell_h
\ee
where we used (I.5.21), (I.7.17) and (I.5.12).
\vskip.2cm

If $X,Y$ form a biangle, so that $s_Y s_X = e$, then
\be
     Q(X,Y)
 &=& \sum_{h \in S} Q^h_{(e) s_X,s_Y} \cdot \ell_h
  = \sum_{h \in S} \delta^e_{s_X s_Y} \, ( \delta^h_{s_X}
     + V^h_{s_X,s_Y} ) \cdot \ell_h
  = X + \sum_{h \in S} V^h{}_{s_X,s_Y} \cdot \ell_h \nonumber \\
 &=& X + \tilde{\V}_X R_{X \ast} Y  \; .
\ee
If $X,Y,Z$ form a triangle, so that $s_Y s_X = s_Z$, we obtain
\be
     Q(X,Y)
 &=& \sum_{h \in S, h_1 \in S_{(1)}} Q^h_{(h_1) s_X,s_Y} \cdot \ell_h
  = \sum_{h \in S, h_1 \in S_{(1)}} \delta^{h_1}_{s_Y s_X} \,
     ( \delta^h_{s_X} - \delta^h_{h_1}
     + V^h{}_{s_X, \ad(s_X)^{-1} s_Y} ) \cdot \ell_h  \nonumber \\
 &=& X + \tilde{\V}_X R_{X \ast} Y - Z \; .
\ee
Finally, for a quadrangle $X,Y,\hat{X},\hat{Y}$ (which satisfies
$s_Y s_X = s_{\hat{Y}} s_{\hat{X}} \not\in S_e$) we find
\be
      Q(X,Y;\hat{X},\hat{Y})
 &:=& Q(X,Y) - Q(\hat{X},\hat{Y})
   = \sum_{h \in S, g \in S_{(2)}}
     Q^h_{(g) s_X, s_Y, s_{\hat{X}}, s_{\hat{Y}} }
     \cdot \ell_h         \nonumber \\
 &=& \sum_{h \in S, g \in S_{(2)}} \delta^g_{s_Y s_X} \,
     ( \delta^h_{s_X} - \delta^h_{s_{\hat{X}}}
     + V^h_{s_X, \ad(s_X)^{-1} s_Y}
     - V^h{}_{ s_{\hat{X}} , \ad(s_{\hat{X}})^{-1} s_{\hat{Y}} } )
     \cdot \ell_h  \nonumber \\
 &=& X + \tilde{\V}_X R_{X\ast}Y - \hat{X}
       - \tilde{\V}_{\hat{X}} R_{\hat{X} \ast} \hat{Y} \, .
\ee
\vskip.2cm

 For arbitrary vector fields $X,Y,Z$ we define the {\em curvature tensor}
\be
   \R(X,Y)(Z)
 = \sum_{h \in S, \, h_1,h_2,h_3 \in S_e} X^{h_1} Y^{h_2} Z^{h_3} \,
   \R^h{}_{h_3,h_1,h_2} \cdot \ell_h
\ee
where the ambiguity in the quadrangle components is fixed by
(\ref{quad_curv_comp}).
If $X,Y,Z$ are basic, we obtain
\be
    \R(X,Y)(Z)
  = \sum_{h \in S} \R^h{}_{s_Z,s_X,s_Y} \cdot \ell_h \; .
\ee
 For further evaluation we need the following expressions,
\be
   \tilde{\V}_{R_{X \ast} Y} Z
 &=& \sum_{h_1} (R_{X \ast} Y)^{h_1} \, \tilde{\V}_{\ell_{h_1}} Z
 = \sum_{h_1,h_2} (R_{X \ast} Y)^{h_1} \, (R^\ast_{h_1} Z^{h_2})
   \, \tilde{\V}_{\ell_{h_1}} \ell_{h_2}  \nonumber \\
 &=& \sum_{h_1,h_2,h} (R_{X \ast} Y)^{h_1} \, (R^\ast_{h_1} Z^{h_2})
   \, V^h{}_{h_1,h_2} \cdot \ell_h
\ee
and
\be
     \tilde{\V}_X \tilde{\V}_{R_{X \ast} Y} Z
 &=& \sum_{h_1} X^{h_1} \, \tilde{\V}_{\ell_{h_1}}( \tilde{\V}_{R_{X \ast} Y} Z )
          \nonumber \\
 &=& \sum_{h,h_1,h_2,h_3} X^{h_1} \, Y^{\ad(h_1)h_2} \,
     (R^\ast_{h_1 h_2} Z^{h_3}) \,
     \left( V_{h_1} \, R^\ast_{h_1} V_{h_2} \right)^h{}_{h_3}
     \cdot \ell_h
\ee
using (I.5.21) and (I.5.12). With the help of
these formulas we obtain
\be
     \R(X,Y)(Z)
 &=& \tilde{\V}_X \tilde{\V}_{R_{X \ast} Y} Z - Z
         \quad \mbox{for a biangle $X,Y$}   \\
     \R(X,Y)(Z)
 &=& ( \tilde{\V}_X \tilde{\V}_{R_{X \ast} Y} - \tilde{\V}_W ) \, R_{W \ast} Z
         \quad \mbox{for a triangle $X,Y,W$}
\ee
and
\be
    \R(X,Y;\hat{X},\hat{Y})(Z)
 &:=& \R(X,Y)(Z) - \R(\hat{X},\hat{Y})(Z) \nonumber \\
 &=& ( \tilde{\V}_X \tilde{\V}_{R_{X \ast} Y}
     -  \tilde{\V}_{\hat{X}} \tilde{\V}_{R_{\hat{X} \ast} \hat{Y}} )
     \, R_{(R_{X \ast} Y) \ast} \, R_{X \ast} Z \\
 & & \mbox{for a quadrangle $X,Y,\hat{X},\hat{Y}$} \; . \nonumber
\ee
\vskip.1cm

The {\em Ricci tensor} defined in (\ref{Riccit1}) can also be
expressed as follows,
\be
    {\it Ric}(X,Y)
 := \sum_{h \in S} \langle \R(\ell_h,Y)(X) , \theta^h \rangle \; .
\ee

\section{Riemannian group lattices admitting a torsion-free compatible
         linear connection}
\label{sec:discrRiem}
\setcounter{equation}{0}
Let $(G,S)$ be a bicovariant group lattice and $(\Omega, \d)$ the
associated differential calculus.
The formalism developed in the previous sections enables us to
carry familiar constructions of continuum differential geometry
over to the discrete differential geometric framework of group
lattices. In particular, we may look for an analog of the
\emph{Levi-Civita connection} of a metric $\mathsf{g}$. This means we should
look for torsion-free linear connections which are compatible with $\mathsf{g}$.
\vskip.1cm

In section~\ref{subsec:compat} a (bicovariant) group lattice supplied
with a metric tensor $\mathsf{g}$ of constant signature has been called
a ``Riemannian group lattice''.
In this section we further demand that it admits a
\emph{torsion-free} metric-compatible linear connection.
Unlike the continuum case, on most group lattices not every metric admits
such a connection. As we shall see below, this condition
indeed places severe restrictions on the components of a metric.
This should not come as a big surprise. In continuum differential
geometry the requirement of a smooth metric on a smooth manifold
guarantees that the metric components at ``neighboring'' points
fit together. On the other hand, given a set of points in a
Euclidean space, for example, and prescribing metric components
at every point, a corresponding embedded digraph does not exist,
in general. This is not the whole story, however. In the case
of a maximal group lattice (complete digraph), which corresponds
to a maximal set $S$, vanishing torsion already determines a
unique linear connection, so that no freedom is left to satisfy
the metric-compatibility conditions for a ``non-trivial'' geometry
(see subsection~\ref{subsec:max}). Reducing $S$ to smaller sets
allows for more freedom in a torsion-free connection and thus
for more solutions of the metric-compatibility conditions.
\vskip.1cm

If a metric-compatible linear connection is found for a given metric,
it is only determined up to transformations
$V_h \mapsto J_h \, V_h$ of the transport matrices with isometry matrices
$J_h$ (see section~\ref{subsec:compat}) with coefficients $J^{h'}{}_{h,h''}$.
Requiring vanishing torsion restricts this freedom, but in general
does not fix it completely.
In the following we elaborate this in more detail. More generally,
we look separately at the consequences of vanishing biangle, triangle
and quadrangle torsion together with the metric-compatibility condition.
In the following, the matrices $J_h$ are always constrained
by the isometry condition (\ref{J-isom}).
\vskip.2cm

\noindent
a) \emph{Vanishing biangle torsion.} The biangle torsion vanishes for
a biangle $h_1 h_2 = e$ (at some lattice site) if and only if
\be
   V^h{}_{h_1,h_2} = - \delta^h_{h_1} \qquad \forall \, h \in S
                     \label{V_bt=0}
\ee
which is $V_{h_1,h_2} = - \ell_{h_1}$. Together with the
metric-compatibility condition (\ref{Vg_compat}), this leads to
\be
   R^\ast_{h_1} \g_{h_2,h} = - \sum_{h' \in S} \g_{h_1,h'} \, V^{h'}{}_{h_1,h}
   \qquad  \forall \, h \in S
\ee
and in particular
\be
    R^\ast_{h_1} \g_{h_2,h_2} = \g_{h_1,h_1} \qquad
    ( h_1 h_2 = e )  \; .     \label{gaa_bt=0}
\ee
It is natural to assign to $\g_{h_1,h_1}$ the interpretation
of the square of the distance from $g$ to $gh_1$. Then the
last formula tells us that this distance is equal to the
reverse distance, i.e. that from $gh_1$ to $g$.
\vskip.1cm
\noindent
{\em Remark.} For making contact with ordinary discrete geometry,
this suggests to demand vanishing biangle torsion. It should be
noticed, however, that (\ref{gaa_bt=0}) does not necessarily
require vanishing biangle torsion
(see sections~\ref{subsec:Z412torsion} and \ref{subsec:Z413torsion}).
Furthermore, in a communication network it is natural to allow
the possibility of assigning different lengths (routing distances)
to a direction and its inverse.
\hfill $\blacksquare$
\vskip.1cm

\noindent
As a consequence of (\ref{V_bt=0}), only transformations of $V_{h_1}$
are allowed with an isometry matrix $J_{h_1}$ subject to
\be
    J^h{}_{h_1,h_1} = \delta^h_{h_1} \qquad  \forall \, h \in S \; .
\ee
This means $J_{h_1} \ell_{h_1} = \ell_{h_1}$, which restricts the
freedom to isometries leaving the vector $V_{h_1,h_2} = -\ell_{h_1}$
invariant. These are rotations (including reflections) about $V_{h_1,h_2}$.
\vskip.2cm

\noindent
b) \emph{Vanishing triangle torsion.} The vanishing of the triangle
torsion for a triangle $h_1 h_2 = h_0$ (at some lattice site) amounts to
\be
    V^h{}_{h_1,h_2} = \delta^h_{h_0} - \delta^h_{h_1} \qquad
     \forall \, h \in S     \label{V_tt=0}
\ee
which is $V_{h_1,h_2} = \ell_{h_0} - \ell_{h_1}$.
Together with (\ref{Vg_compat}) this implies
\be
   R^\ast_{h_1} \g_{h_2,h} = \sum_{h' \in S} (\g_{h_0,h'} - \g_{h_2,h'})
   \, V^{h'}{}_{h_1,h}
   \qquad  \forall \, h \in S
\ee
and in particular
\be
    R^\ast_{h_1} \g_{h_2,h_2}
  = \g_{h_1,h_1} + \g_{h_0,h_0} - 2 \, \g_{h_0,h_1}
    \qquad ( h_1 h_2 = h_0 )  \; .
\ee
Using the standard interpretation of the metric components, this is
a well-known law of Euclidean geometry, the cosine law of triangles.
Hence, the requirement of a metric-compatible and triangle-torsion-free
linear connection restricts the metric in such a way that triangles are
always flat. If triangle torsion is admitted, however, then it is possible
to curve a triangle in such a way, for example, that the parallel
transport is that of a spherical triangle, see section~\ref{subsec:Z3t}.
\vskip.1cm
\noindent
(\ref{V_tt=0}) restricts the freedom of isometries in the transport matrices by
\be
  J^h{}_{h_1,h_0} - J^h{}_{h_1,h_1} = \delta^h_{h_0} - \delta^h_{h_1}
     \qquad \forall \, h \in S
\ee
which is $J_{h_1}(\ell_{h_0} - \ell_{h_1}) = \ell_{h_0} - \ell_{h_1}$.
Hence $J_{h_1}$ corresponds to a ``rotation'' which leaves
the vector $V_{h_1,h_2} = \ell_{h_0} - \ell_{h_1}$ fixed.
\vskip.2cm

\noindent
c) \emph{Vanishing quadrangle torsion.} The vanishing of the quadrangle
torsion associated with a quadrangle
$h_1 h_2 = \hat{h}_1 \hat{h}_2 = g \not\in S_e$ (at some lattice site) means
\be
     V^h{}_{h_1,h_2} + \dl^h_{h_1}
   = V^h{}_{\hat{h}_1,\hat{h}_2} + \dl^h_{\hat{h}_1}
     \qquad \forall \, h \in S
\ee
and thus $V_{h_1,h_2} - V_{\hat{h_1},\hat{h}_2} = \ell_{\hat{h}_1} - \ell_{h_1}$.
Together with the metric-compatibility condition this imposes restrictions
on the metric components. In particular, for a positive definite metric
the triangle inequalities lead to
\be
   \Big| \| V_{h_1,h_2} \| - \| V_{\hat{h}_1,\hat{h}_2} \| \Big|
   \leq \| \ell_{h_1}-\ell_{\hat{h}_1} \|
   \leq \| V_{h_1,h_2} \| + \| V_{\hat{h}_1,\hat{h}_2} \|
\ee
where $\| V_{h_1,h_2} \| = \sqrt{g(V_{h_1,h_2},V_{h_1,h_2})}$.
Using (\ref{Vg_compat}), this restricts the metric components as follows,
\be
  \Big| \sqrt{R^\ast_{h_1}\g_{h_2,h_2}}-\sqrt{R^\ast_{\hat{h}_1}
  \g_{\hat{h}_2,\hat{h}_2}} \Big|
  \leq \sqrt{\g_{h_1,h_1}+\g_{\hat{h}_1,\hat{h}_1}-2 \, \g_{h_1,\hat{h}_1}}
  \leq \sqrt{R^\ast_{h_1}\g_{h_2,h_2}}+\sqrt{R^\ast_{\hat{h}_1}
  \g_{\hat{h}_2,\hat{h}_2}}  \, . \quad   \label{0_quadr_torsion_ineq}
\ee

The isometries $J_h$ have to satisfy the equation
\be
  \sum_{h' \in S} ( J^h{}_{h_1,h'} - J^h{}_{\hat{h}_1,h'} ) \, V^{h'}{}_{h_1,h_2}
  =  J^h{}_{\hat{h}_1,h_1} - \dl^h_{h_1} - J^h{}_{\hat{h}_1,\hat{h}_1}
     + \dl^h_{\hat{h}_1}
     \qquad \forall \, h \in S    \label{J-quadr}
\ee
which is $J_{h_1} V_{h_1,h_2} - J_{\hat{h}_1} V_{\hat{h}_1,\hat{h}_2}
= \ell_{\hat{h}_1} - \ell_{h_1}$. In particular, a rotation
which leaves $V_{h_1,h_2}$ fixed, so that $J_{h_1} V_{h_1,h_2} = V_{h_1,h_2}$,
together with a rotation which leaves $V_{\hat{h}_1,\hat{h}_2}$ fixed, so that
$J_{\hat{h}_1} V_{\hat{h}_1,\hat{h}_2} = V_{\hat{h}_1,\hat{h}_2}$,
preserves the quadrangle and thus solves the above constraint.
Another possibility is given by combined rotations $J_{h_1}$ and
$J_{\hat{h}_1}$ which leave the vector
$V_{h_1,h_2} - V_{\hat{h}_1,\hat{h}_2}$ and thus $\ell_{\hat{h}_1}-\ell_{h_1}$
fixed, so that
$J_{h_1}(\ell_{h_1} - \ell_{\hat{h}_1}) = \ell_{h_1} - \ell_{\hat{h}_1}$
and
$J_{\hat{h}_1}(\ell_{h_1} - \ell_{\hat{h}_1}) = \ell_{h_1} - \ell_{\hat{h}_1}$.
\vskip.2cm

The following subsections provide several examples of Riemannian group
lattices which admit torsion-free linear connections. In the discussions
we make use of the fact that a linear connection determines a
tangent space picture of the group lattice, as described in
section~\ref{subsec:mc-conn-interpr}.

\subsection{Maximal group lattices}
\label{subsec:max}
A group lattice $(G,S)$ with $S = G \setminus \{e\}$ is called {\em maximal}.
It is bicovariant and carries the universal differential calculus.
In this case there are only biangles and triangles, but no quadrangles.
The condition of vanishing torsion then determines a unique
linear connection which is given by
\be
   V^h{}_{h_1,h_2} = \left \{ \begin{array}{r@{\quad\mbox{if}\quad}r}
     - \delta^h_{h_1} & h_1 h_2 = e \\
     \delta^h_{h_0} - \delta^h_{h_1} & h_0 := h_1 h_2 \neq e
     \end{array} \right.
\ee
and thus constant. This implies $V_h V_{h^{-1}} = I$ and
$V_{h_1} V_{h_2} = V_{h_0}$ if $h_1 h_2 = h_0$. As a consequence,
the curvature of the connection vanishes. \cite{rem:conn_udc}
\vskip.1cm

The metric compatibility condition evaluated for this connection
becomes
\be
  R_h^\ast \g_{h_1,h_2} = \left \{ \begin{array}{rcr}
     \g_{h,h} - \g_{h,hh_2} - \g_{hh_1,h} + \g_{hh_1,hh_2} &
      & h h_1 \neq e, \, h h_2 \neq e \\
     \g_{h,h} - \g_{h,hh_2} & \quad\mbox{if}\quad & h h_1 = e, \, h h_2 \neq e \\
     \g_{h,h} & & h_1 = h_2, \, h h_1 = e
     \end{array} \right.
\ee
\vskip.2cm

\begin{example}
\label{ex:Z3T=0}
Let $G$ be $\mathbb{Z}_3$, the cyclic group consisting of the three
elements $0,1,2$ with addition modulo 3 as the group composition.
We choose the group lattice determined by $S = \{ 1,2 \}$ which is
the complete digraph with three vertices. There are two biangles,
$1+2 = 0 = 2+1$ (modulo $3$), and two triangles, $1+1=2$ and $2+2=1$
(modulo $3$). The unique torsion-free linear connection is
determined by the two matrices
\be
  V_1 = \left(\begin{array}{cc} -1 & -1 \\
                                 1 & 0
              \end{array} \right) \, , \qquad
  V_2 = \left(\begin{array}{cc} 0 &  1 \\
                               -1 & -1
              \end{array} \right)  \; .
\ee
A metric is given by
\be
    \g = \left(\begin{array}{cc} a & b \\
                                       b & c
               \end{array} \right)   \label{metric_abc}
\ee
with functions $a,b,c$ and the compatibility condition with
the above connection reduces to
\be
  R_1^\ast \left(\begin{array}{cc} a & b \\
                                   b & c
                 \end{array} \right)
  = \left(\begin{array}{cc} a-2b+c & a-b \\
                               a-b & a
          \end{array} \right) \; .
\ee
This means that one can specify arbitrary values of the metric functions
$a,b,c$ at {\em one} point. The metric at the other points is then
determined by the last equation and the resulting metric on $\mathbb{Z}_3$
is compatible with the above torsion-free connection. Assigning the usual
interpretation in terms of Euclidean distances and angles to the
metric components, one recovers the rules of Euclidean trigonometry.
In particular, in case of a constant metric,
the compatibility condition restricts $\g$ to
\be
           \g = a \left(\begin{array}{cc} 1 & 1/2 \\
                                        1/2 & 1
                        \end{array} \right)
\ee
with a constant $a$. This expresses metric properties of a regular
Euclidean triangle.
The parallel transport determined by the torsion-free connection
coincides with that of the Euclidean plane.
Indeed, from (\ref{tVell-ell}) we infer
\begin{center}
\begin{tabular}{cc}
\begin{tabular}{c|ccc}
    & at $k+1$ mod $3$ & & at $k$ mod $3$ \\
  \hline
  $\tilde{\V}_{\ell_1} :$ & $\begin{array}{c} \ell_1 \\ \ell_2 \end{array}$ &
  $\mapsto$ & $\begin{array}{c} \ell_2 - \ell_1 \\ - \ell_1 \end{array}$
\end{tabular}  &
\begin{tabular}{c|ccc}
    & at $k+2$ mod $3$ & & at $k$ mod $3$ \\
  \hline
  $\tilde{\V}_{\ell_2} :$ & $\begin{array}{c} \ell_1 \\ \ell_2 \end{array}$ &
  $\mapsto$ & $\begin{array}{c} - \ell_2 \\ \ell_1 - \ell_2 \end{array}$
\end{tabular}
\end{tabular}
\end{center}
which maps the Riemannian group lattice isometrically onto a Euclidean
triangle in the tangent space at a site.
\end{example}
\vskip.2cm

\begin{example}
\label{ex:Z4T=0}
Let $G = \mathbb{Z}_4$ and $S = \{ 1,2,3 \}$. The corresponding torsion-free
linear connection is then given by
\be
  V_1 = \left( \begin{array}{rrr}
   -1 & -1 & -1 \\ 1 & 0 & 0 \\ 0 & 1 & 0 \end{array} \right) \, , \quad
  V_2 = \left( \begin{array}{rrr}
   0 & 0 & 1 \\ -1 & -1 & -1 \\ 1 & 0 & 0 \end{array} \right) \, , \quad
  V_3 = \left( \begin{array}{rrr}
   0 & 1 & 0 \\ 0 & 0 & 1 \\ -1 & -1 & -1 \end{array} \right) \, .
\ee
Assuming the metric to be constant, the compatibility
condition restricts it to the form
\be
  \g = \left( \begin{array}{ccc}
       a     & b      & a - b \\
       b     & 2 \, b & b     \\
       a - b & b      & a
       \end{array} \right)  \; .
\ee
For $b = a/2$ we recover the geometry of a regular tetrahedron in Euclidean
$\mathbb{R}^3$. Since we deal with a three-dimensional Riemannian group
lattice, we are actually describing the tetrahedron volume.
Furthermore, in the limit $b \to a$ the above geometry tends to that
of a quadrate in the Euclidean plane where the vector associated with
$2 \in S$ corresponds to the diagonal. Accordingly, in this limit the
determinant of $\g$ vanishes, so that $\g$ no longer defines a metric
according to our definition in section~\ref{subsec:compat}.
\end{example}

\subsection{Two-dimensional Riemannian group lattices}
\label{subsec:2dRgl}
Let $G$ be a discrete group, $S = \{ a,b \}$ a subset
consisting of two different elements of $G$ which generate $G$
such that $(G,S)$ is a bicovariant group lattice. Then either
$aba^{-1}=a$, which contradicts $a \neq b$, or $a b a^{-1} = b$
which is $ab = ba$. Hence, bicovariance requires that $G$ is
Abelian. By a fundamental theorem, every finite Abelian group is
isomorphic to a direct product of cyclic groups of prime power
order.
\vskip.1cm

The following examples in particular demonstrate that, for a given
metric on a group lattice, there may not exist a metric-compatible
linear connection with vanishing torsion, i.e. a Levi-Civita connection.
Moreover, in contrast to ordinary continuum differential
geometry, if such a connection exists, then is not unique.

\subsubsection{$\mathbb{Z}_4$ lattices}
\label{subsec:Z4}
a) Let $G = \mathbb{Z}_4$ and $S = \{ 1,2 \}$. There is one biangle,
$2+2 = 0$ (modulo $4$), one triangle, $1+1 = 2$, and one quadrangle,
$1+2 = 3 = 2+1$, which implies the 2-form relation
$\theta^1 \cap \theta^2 = - \theta^2 \cap \theta^1$.
Vanishing torsion restricts the matrices $V_i$ of the linear
connection to
\be
  V_1 = \left(\begin{array}{cc} -1 & p \\ 1 & 1 + q \end{array} \right)
        \, , \qquad
  V_2 = \left(\begin{array}{cc} 1+p & 0 \\ q & -1 \end{array} \right)
\ee
with arbitrary functions $p$ and $q$. For a metric of the form
(\ref{metric_abc}), the metric-compatibility condition
$R_1^\ast \g = V_1^T \g V_1$ reads
\be
  R_1^\ast a &=& a - 2 \, b + c  \nonumber \\
  R_1^\ast b &=& p \, (b-a) + (1+q)(c-b) \nonumber \\
  R_1^\ast c &=& p^2 \, a + 2 \, p \, (1+q) \, b + (1+q)^2 \, c  \; .
\ee
With the help of $R_2^\ast = (R_1^\ast)^2$, the second condition
$R_2^\ast \g = V_2^T \g V_2$ leads to
\be
    V_1 \, (R_1^\ast V_1) \, V_2^{-1} = J
\ee
where $J$ is an arbitrary element of the isometry group of the metric
(at each site of the group lattice). A lengthy computation,
aided by computer algebra, reveals that
\emph{every Levi-Civita connection on $(\mathbb{Z}_4, \{ 1,2 \})$ is flat,
i.e. its curvature vanishes}. \cite{Z4_math}
The integrability condition (\ref{integr_mc_triangle}) then enforces $J=I$
so that $V_2 = V_1 \, R_1^\ast V_1$.
As a consequence, we obtain the following representation
of $\mathbb{Z}_4$:
\be
    R_1^\ast p = - {p \over 1+p+q} \, , \qquad
    R_1^\ast q = - {2+p+q \over 1+p+q}  \; .
\ee
This implies $R_2^\ast p = -p/(1+p)$ and $R_2^\ast q = q/(1+p)$
and thus $(R_1^\ast)^4 p = (R_2^\ast)^2 p = p$,
$(R_1^\ast)^4 q = (R_2^\ast)^2 q = q$.

Excluding special values of $q(0)$ and $p(0)$,
the geometries with a Levi-Civita connection are given by
\be
    a(1) &=& a(0) - 2 \, b(0) + c(0) \nonumber \\
    b(1) &=& - p(0) \, a(0) + [p(0) - 1 -q(0)] \, b(0) + [1+q(0)] \, c(0)
                \nonumber \\
    c(1) &=& p(0)^2 \, a(0) + 2 \, p(0) \, [1+q(0)] \, b(0) + [1+q(0)]^2 \, c(0)
                \nonumber \\
    a(2) &=& [1+p(0)]^2 \, a(0) + 2 \, q(0) \, [1+p(0)] \, b(0) + q(0)^2 \, c(0)
                \nonumber \\
    b(2) &=& - [1+p(0)] \, b(0) - q(0) \, c(0) \nonumber \\
    c(2) &=& c(0) \nonumber \\
    a(3) &=& [1+p(0)]^2 \, a(0) + 2 \, [1+p(0)] \, [1+q(0)] \, b(0)
             + [1+q(0)]^2 \, c(0)              \nonumber \\
    b(3) &=& p(0) \, [1+p(0)] \, a(0)
         + [1+2 \, p(0)] \, [1+q(0)] \, b(0)
         + [1+q(0)]^2 \, c(0)            \nonumber \\
    c(3) &=& p(0)^2 \, a(0) + 2 \, p(0) \, [1+q(0)] \, b(0) + [1+q(0)]^2 \, c(0)
\ee
and
\be
  q(1) &=& -[2+p(0)+q(0)]/[1+p(0)+q(0)] \, , \quad
  p(1) = - p(0)/[1+p(0)+q(0)] \nonumber  \\
  q(2) &=& q(0)/[1+p(0)] \, , \quad
  p(2) = - p(0)/[1+p(0)] \nonumber \\
  q(3) &=& -[2+p(0)+q(0)]/[1+q(0)] \, , \quad
  p(3) = p(0)/[1+q(0)] \; .
\ee
\vskip.2cm

\noindent
b) Let $G = \mathbb{Z}_4$ again, but now we choose $S = \{ 1,3 \}$.
In this case, there are two biangles, $1+3 = 0 = 3+1$ (modulo $4$),
no triangle and a quadrangle corresponding to $1+1 = 2 = 3+3$
(modulo $4$). The latter leads to the 2-form relation
$\theta^1 \cap \theta^1 + \theta^3 \cap \theta^3 = 0$.
The condition of vanishing torsion imposes the following restrictions
on a linear connection,
\be
  V_1 = \left(\begin{array}{cc} u & -1 \\ 1+v & 0 \end{array} \right)
        \, , \qquad
  V_3 = \left(\begin{array}{cc} 0 & 1+u \\ -1 & v \end{array} \right)
\ee
with arbitrary functions $u$ and $v$. For a metric of the form
(\ref{metric_abc}) the compatibility condition
$R_1^\ast \g = V_1^T \g V_1$ reads
\be
   R_1^\ast a = u^2 \, a + 2 u (1+v) \, b + (1+v)^2 \, c \, , \quad
   R_1^\ast b = - u \, a - (1+v) \, b  \, , \quad
   R_1^\ast c = a
\ee
and, with the help of $R_3^\ast = (R_1^\ast)^3$, the second
metric-compatibility condition
$R_3^\ast \g = V_3^T \g V_3$ leads to
\be
    V_1 \, (R_1^\ast V_1) [(R_1^\ast)^2 V_1] \, V_3^{-1} = J
\ee
where $J$ is an element of the isometry group of the
metric. Further exploration with the help of computer algebra
shows that \emph{every Levi-Civita connection on $(\mathbb{Z}_4, \{ 1,3 \})$
has vanishing curvature}.
\vskip.2cm

Since the only metric-compatible torsion-free linear connections
on the above $\mathbb{Z}_4$ lattices have vanishing curvature,
via backward parallel transport they are isometrically mapped
to a closed lattice in $\mathbb{R}^2$ which represents the tangent
space at a site. In particular, this means that we cannot
model something like a tetrahedron surface in this way.
To supply the $\mathbb{Z}_4$ group lattices with
non-vanishing curvature is only possible if the condition
of vanishing torsion is dropped (see sections~\ref{subsec:Z412torsion}
and \ref{subsec:Z413torsion}).

\subsubsection{$\mathbb{Z}^2$ lattices}
Let us consider the group lattice $(\mathbb{Z}^2, \{ \hat{1}, \hat{2} \})$
where $\hat{1} := (1,0)$ and $\hat{2} := (0,1)$. It has no biangles or
triangles, but a quadrangle corresponding to
$\hat{1} + \hat{2} = \hat{2} + \hat{1}$.
The condition of vanishing torsion restricts the parallel transport
matrices $V_i := V_{\hat{i}}$ to
\be
  V_1 = \left(\begin{array}{cc} p & u \\ q & 1+v \end{array} \right)
        \, , \qquad
  V_2 = \left(\begin{array}{cc} 1+u & r \\ v & s \end{array} \right)
     \label{Z^2_Vs}
\ee
with arbitrary functions $p,q,r,s,u,v$. The following example demonstrates
that there are torsion-free and metric-compatible parallel transports with
\emph{non}-vanishing curvature even on a two-dimensional lattice carrying the
metric properties of a regular quadratic lattice in Euclidean $\mathbb{R}^2$.
\vskip.2cm

\begin{example}
\label{ex:Z2-reglatt}
Let us choose the metric to be
\be
   \g = \left(\begin{array}{cc} 1 & 0 \\ 0 & 1 \end{array} \right)
\ee
at all sites. The metric-compatibility condition for the above torsion-free
linear connection then leads to the following two classes of solutions.
The first class is given by
\be
  V_1 = \left(\begin{array}{cc} \epsilon_1 & 0 \\ 0 & 1 \end{array} \right)
        \, , \qquad
  V_2 = \left(\begin{array}{cc} 1 & 0 \\ 0 & \epsilon_2 \end{array} \right)
\ee
with functions $\epsilon_i$ with values in $\{ \pm 1 \}$. The curvature only
vanishes if $\epsilon_1$ and $\epsilon_2$ are constant in the
$\hat{1}$ and $\hat{2}$ direction, respectively.

If the curvature vanishes, the (backward) parallel transport does not depend
on the path in the lattice, see (\ref{curv&ptransport}). It can thus be used
to map the whole group lattice into the tangent space at one point, which is
isomorphic to $\mathbb{R}^2$ in the case under consideration.
Let us choose the lattice point $(0,0)$.
The tangent vectors $\ell_h$ at this site may then be identified with the
vectors $\mathbf{u}_1$ and $\mathbf{u}_2$ pointing from $(0,0)$ to $(1,0)$
and $(0,1)$, respectively, in $\mathbb{R}^2$.
Then $\ell_{\hat{1}}$ at the group lattice site $(1,0)$ is mapped to the vector
$\mathbf{V}_{1,1}$ which we attach at the tip of $\mathbf{u}_1$ in
$\mathbb{R}^2$ according to the prescription of section~\ref{subsec:mc-conn-interpr}.
If $\epsilon_1 = -1$ this vector points into the ``wrong direction'', i.e.
its tip coincides with $(0,0)$. This means that the resulting lattice in
$\mathbb{R}^2$ gets \emph{folded}. Similarly, if $\epsilon_2 = -1$ the lattice
gets folded in the other direction.

A particular solution is given by $V_h = I$, the unit matrix, at
all sites. It corresponds to the ordinary Euclidean parallel transport.
This solution certainly has a nice continuum limit.
Introducing a lattice spacing parameter, we may write
$V_h = I + \kappa \, \Gamma_h + O(\kappa^2)$.
Some of the other solutions $V_h$ given above have negative determinant
at some sites and cause folding in the sense described above. They are
related to the above solution at those sites by an isometry $J_h$
with determinant $-1$. As a consequence, they cannot have a continuum limit.
The requirement of a continuum limit may thus distinguish
a certain connection and eliminate connections with folding.

The second class of solutions is given by
\be
  V_1 = \left(\begin{array}{cc} 0 & -1 \\ \epsilon_1 & 0 \end{array} \right)
        \, , \qquad
  V_2 = \left(\begin{array}{cc} 0 & \epsilon_2 \\ -1 & 0 \end{array} \right)
        \; .
\ee
The curvature only vanishes if at all sites
$R_{\hat{2}}^\ast \epsilon_1 = \epsilon_2$
and $R_{\hat{1}}^\ast \epsilon_2 = \epsilon_1$.
An orientation-preserving connection is obtained if
$\epsilon_1=\epsilon_2=1$. The corresponding transports in
the two directions act with rotations.
\end{example}

\begin{figure}
\begin{center}
\includegraphics[scale=.6]{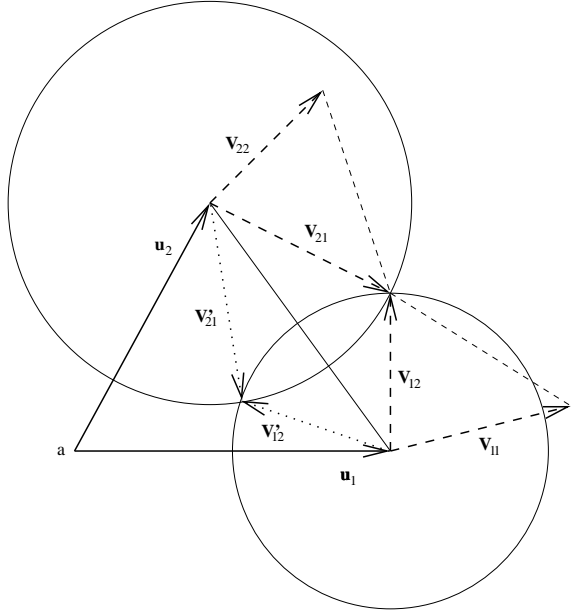}
\caption{Levi-Civita connections on a $\mathbb{Z}^2$ group lattice exist
if and only if at each lattice site the circle with radius $|\mathbf{V}_{12}|$
around the tip of $\mathbf{u}_1$ intersects the circle with radius $|\mathbf{V}_{21}|$
around the tip of $\mathbf{u}_2$.}\label{fig:concon}
\end{center}
\end{figure}

\begin{figure}
\begin{center}
\includegraphics[scale=.6]{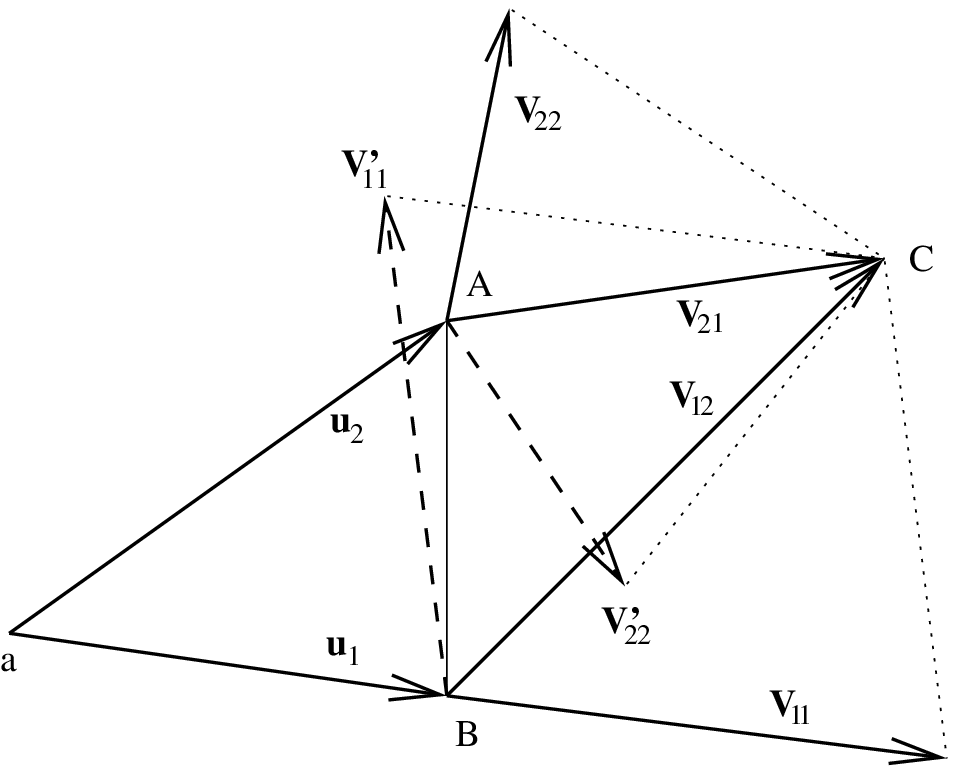}
\qquad
\includegraphics[scale=.6]{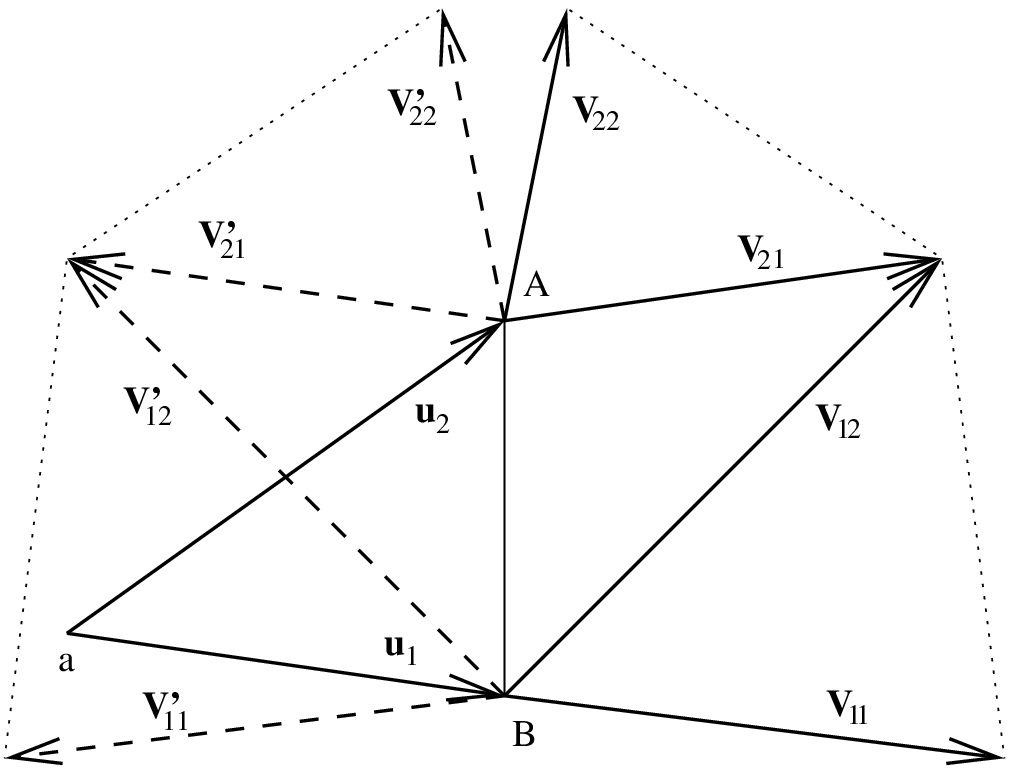}
\caption{The vectors $\mathbf{V}_{ij}$ and $\mathbf{V}'_{ij}$ result from
torsion-free metric-compatible (backward) parallel transports which differ
by a reflection about some axis.}\label{fig:levi}
\end{center}
\end{figure}

There are metrics (with constant signature) on
$(\mathbb{Z}^2, \{ \hat{1}, \hat{2} \})$ which do \emph{not}
admit a Levi-Civita connection, although the constraints
are by far not as stringent as in our previous examples.
Counterexamples are easily constructed. A geometric condition
for the existence of a Levi-Civita connection is given by
(\ref{0_quadr_torsion_ineq}) in the case of a positive definite
metric. Let us recall its origin in the case under consideration.
The tangent space at a site $a$ is isomorphic to $\mathbb{R}^2$
with the Euclidean inner product of vectors
(see section~\ref{subsec:mc-conn-interpr}).
The tangent vectors $\ell_{\hat{i}}$ are then represented by
vectors $\mathbf{u}_i \in \mathbb{R}^2$, $i=1,2$,
such that $\mathbf{u}_i \cdot \mathbf{u}_j = \g_{ij}(a)$
where $\g_{ij} := \g_{\hi,\hj}$.
The parallel transport $\tilde{\V}_{\ell_\hi}$ maps the tangent
space at the site $a+\hi$ into the tangent space at $a$.
Metric-compatibility of the connection means
\be
   \mathbf{V}_{ij} \cdot \mathbf{V}_{ik} = \g_{j k}(a+\hi)
       \label{Z2-mcompat}
\ee
where $\mathbf{V}_{ij}$ represents $\tilde{\V}_{\ell_\hi} \ell_\hj$
at the site $a$. If the connection is (quadrangle) torsion free,
then adjacent quadrangles are preserved by the backward parallel
transport, so that
$\mathbf{u}_i + \mathbf{V}_{ij} = \mathbf{u}_j + \mathbf{V}_{ji}$.
The last equation has solutions if and only if
$\big| |\mathbf{V}_{12}|-|\mathbf{V}_{21}| \big| \leq
|\mathbf{u}_2 - \mathbf{u}_1| \leq |\mathbf{V}_{12}|+|\mathbf{V}_{21}|$
where $|\mathbf{V}_{12}|$ denotes the Euclidean norm of $\mathbf{V}_{12}$
in $\mathbb{R}^2$. This is illustrated in Fig.~\ref{fig:concon}.
Using (\ref{Z2-mcompat}), this condition is expressed as
\be
  \Big| \sqrt{\g_{22}(a+\hat{1})} - \sqrt{\g_{11}(a+\hat{2})} \Big| &\leq&
  \sqrt{\g_{11}(a) + \g_{22}(a) - 2 \, \g_{12}(a)} \nonumber \\
  &\leq& \sqrt{\g_{22}(a+\hat{1})} + \sqrt{\g_{11}(a+\hat{2})}
     \label{Z2-LCcondition}
\ee
in terms of the metric at the sites $a$, $a+\hat{1}$ and $a+\hat{2}$ (see
also (\ref{0_quadr_torsion_ineq})).
If this condition is not fulfilled, a Levi-Civita connection does not exist.
If the condition is satisfied, a Levi-Civita connection exists, but
it is not unique. Even if equality holds in the last part of (\ref{Z2-LCcondition}),
so that the circles in Fig.~\ref{fig:concon} have exactly one point in common,
we still have the freedom to choose $\mathbf{V}_{11}$ and $\mathbf{V}_{22}$
in two possible ways, as illustrated in Fig.~\ref{fig:levi}.
\vskip.1cm

The freedom in the parallel transport left by the conditions of
metric-compatibility and vanishing torsion is a freedom of reflections
about some axes. In Fig.~\ref{fig:levi} it shows up as reflections about
the three axes $AC$, $BC$ and $AB$. In section~\ref{subsec:Zn-LC-freedom}
we show that reflections about $AB$ and $BC$ and their composition
comprise the whole freedom left for $V_1$ and $V_2$ by the conditions
of vanishing torsion and metric-compatibility.
Such reflections lead to folding of the tangent space lattice
obtained by backward parallel transport of the group lattice to the
tangent space at $a$. Moreover, the orientation of some of the
frames at $a$ obtained by backward parallel transport of frames
of basic tangent vectors at $a+\hi$, $i=1,2$, gets changed.
This can be excluded by demanding that $\det V_i>0$.
But we should also require that the dyad $(\mathbf{V}_{21},\mathbf{V}_{12})$
has positive orientation, which is necessary in order to avoid
reflections about the axis $AB$. This amounts to $V^1{}_{12} + V^2{}_{12} > 0$.
In higher dimensions, the determination of the ambiguities in the
Levi-Civita connections and their reduction appears to be a difficult
task (see also section~\ref{subsec:Zn-LC-freedom}).
\vskip.1cm

Whereas torsion is a first order quantity, curvature is of second order
since it expresses features of the geometry determined by the composition
of \emph{two} (backward) parallel transports. In the case under consideration,
the components of the curvature tensor are given in matrix form by
$\R_{ij} := V_i \, R^\ast_\hi V_j - V_j \, R^\ast_\hj V_i$.
\begin{figure}
\begin{center}
\includegraphics[scale=.7]{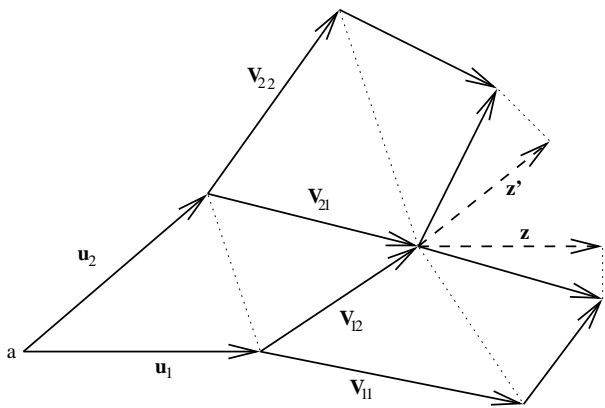}
\caption{The familiar effect of curvature: backward parallel transport
along different paths results in different vectors in the tangent
space at a point.}\label{fig:curvquad}
\end{center}
\end{figure}
In Fig.~\ref{fig:curvquad} the vector $\mathbf{z}$ represents
$\tilde{\V}_{\ell_1} \tilde{\V}_{\ell_2} Z$ where
$Z=\sum_i Z^i\cdot\ell_\hi$. Hence
\be
 \mathbf{z} = \sum_{i,j} \mathbf{u}_i \, [V_1(a) V_2(a+\hat{1})]^i{}_j
              \, Z^j(a+\hat{1}+\hat{2}) \; .
\ee
The vector $\mathbf{z}'$ represents $\tilde{\V}_{\ell_2} \tilde{\V}_{\ell_1} Z$,
so that
\be
  \mathbf{z}{\,}' = \sum_{i,j} \mathbf{u}_i \, [V_2(a) V_1(a+\hat{2})]^i{}_j
                    \, Z^j(a+\hat{2}+\hat{1}) \; .
\ee
The difference gives a measure of the curvature at $a$:
\be
 \mathbf{z} - \mathbf{z}{\,}' = \sum_{i,j} \mathbf{u}_i \, \R^i{}_{j12}(a)
                                \, Z^j(a+\hat{1}+\hat{2})
\ee
(see also (\ref{curv&ptransport})).
\vskip.1cm

If the torsion vanishes at a site $a$, one can draw an isometric picture
of the geometry in the tangent space at $a$ to first order. This represents
the site $a$, its first order neighbors $a+\hi$, and the basic tangent
vectors at these sites while preserving the metric properties
at all these sites and preserving biangles, triangles and quadrangles
at $a$. If moreover the curvature vanishes at $a$, then we can draw an
isometric picture to second order.

\subsection{The freedom in the choice of a Levi-Civita connection
            on a hypercubic $\mathbb{Z}^n$ lattice}
\label{subsec:Zn-LC-freedom}
We already learned that, in general, there is no Levi-Civita connection
for a given metric on a group lattice. If such a connection exists,
it need not be unique. The corresponding freedom will be explored
in this section for the case of hypercubic $\mathbb{Z}^n$ lattices
given by $G = \mathbb{Z}^n$ and $S = \{ \hi \, | \, i = 1, \ldots,n \}$,
where $\hi := (0,\ldots,1,\ldots,0)$ with the 1 in the $i$th position.
We consider only positive definite metrics and choose the standard
inner product $(\mathbf{u},\mathbf{v}) = \mathbf{u} \cdot \mathbf{v}$
for $\mathbf{u},\mathbf{v} \in \mathbb{R}^n$
(cf. section~\ref{subsec:mc-conn-interpr}).
\vskip.1cm

In the case of a hypercubic group lattice
the condition of vanishing torsion can be expressed as
\be
  \mathbf{u}_i + \mathbf{V}_{ij} = \mathbf{u}_j + \mathbf{V}_{ji} \, .
    \label{tf}
\ee
Together with the metric-compatibility, this determines a Levi-Civita
connection up to isometries $J_i$ which preserve the above conditions, i.e.
\be
  \mathbf{u}_i + J_i(\mathbf{V}_{ij}) = \mathbf{u}_j + J_j(\mathbf{V}_{ji})
    \label{jtf}
\ee
(see also (\ref{JhV-vector})).
Subtracting (\ref{tf}) from (\ref{jtf}), we find
\be
    \mathbf{A}_{ij} = \mathbf{A}_{ji} \quad \mbox{where} \quad
  \mathbf{A}_{ij} := J_i(\mathbf{V}_{ij}) - \mathbf{V}_{ij} \, .
     \label{bas}
\ee
Using the isometry condition (\ref{JhV-isom}) and the last equation, we obtain
\be
    \mathbf{V}_{ij} \cdot \mathbf{V}_{ij}
  = J_i(\mathbf{V}_{ij}) \cdot J_i(\mathbf{V}_{ij})
  = \mathbf{V}_{ij} \cdot \mathbf{V}_{ij} + \mathbf{A}_{ij}
    \cdot ( \mathbf{V}_{ij} + \mathbf{V}_{ij} )
\ee
so that
\be
  \mathbf{A}_{ij} \cdot (\mathbf{A}_{ij} + 2 \, \mathbf{V}_{ij}) = 0
          \label{orth1}
\ee
and because of (\ref{bas}) also
\be
  \mathbf{A}_{ij} \cdot (\mathbf{A}_{ij} + 2 \, \mathbf{V}_{ji}) = 0 \; .
     \label{orth2}
\ee
Subtracting the last two equations and using (\ref{tf}) leads to
\be
  \mathbf{A}_{ij} \cdot (\mathbf{u}_j - \mathbf{u}_i) = 0 \; .
        \label{na}
\ee
For $i \neq j$ and if $\mathbf{A}_{ij} \neq 0$, we set
$\mathbf{A}_{ij} = \alpha_{ij} \, \mathbf{a}_{ij}$
with a unit vector $\mathbf{a}_{ij}$ orthogonal to $\mathbf{u}_j - \mathbf{u}_i$.
 From (\ref{orth1}) we then obtain
$\alpha_{ij} = - 2 \, \mathbf{a}_{ij} \cdot \mathbf{V}_{ij} = 0$,
so that
\be
  \mathbf{A}_{ij} = - 2 \, (\mathbf{a}_{ij} \cdot \mathbf{V}_{ij}) \, \mathbf{a}_{ij}
\ee
and thus
\be
 J_i(\mathbf{V}_{ij}) = \mathbf{V}_{ij} - 2 \,
     (\mathbf{a}_{ij} \cdot \mathbf{V}_{ij}) \, \mathbf{a}_{ij} \, , \quad
 J_j(\mathbf{V}_{ji}) = \mathbf{V}_{ji} - 2 \,
     (\mathbf{a}_{ij} \cdot \mathbf{V}_{ji}) \, \mathbf{a}_{ij} \; .
\ee
As a consequence, the effect of $J_i$ on $\mathbf{V}_{ij}$ is that of a
reflection with respect to the hyperplane orthogonal to $\mathbf{a}_{ij}$
(which in turn is orthogonal to $\mathbf{u}_j - \mathbf{u}_i$).
If $\mathbf{A}_{ij} \neq 0$ for all $j \neq i$, then $J_i$ for a fixed $i$
reflects all the $n-1$ vectors $\mathbf{V}_{ij}$, $j \neq i$, with respect to the
respective hyperplane (orthogonal to $\mathbf{a}_{ij}$).
Of course, we still have to respect the remaining conditions which arise
from the isometry conditions (\ref{JhV-isom}), i.e.
$J_i(\mathbf{V}_{ik}) \cdot J_i(\mathbf{V}_{il})
  = \mathbf{V}_{ik} \cdot \mathbf{V}_{il}$.
\vskip.1cm

Let us look at the two-dimensional case.
If $\mathbf{A}_{12} = 0$, we have $J_1(\mathbf{V}_{12}) = \mathbf{V}_{12}$
and $J_2(\mathbf{V}_{21}) = \mathbf{V}_{21}$,
so $J_1$ and $J_2$ are reduced to reflections about $\mathbf{V}_{12}$ and
$\mathbf{V}_{21}$, respectively.
If $\mathbf{A}_{12} \neq 0$, then we have
$J_1(\mathbf{V}_{12}) = \mathbf{V}_{12}
  - 2 \, (\mathbf{a}_{12} \cdot \mathbf{V}_{12}) \, \mathbf{a}_{12}$,
$J_2(\mathbf{V}_{21}) = \mathbf{V}_{21}
  - 2 \, (\mathbf{a}_{12} \cdot \mathbf{V}_{21}) \, \mathbf{a}_{12}$.
The effect of both is a reflection about the axis along $\mathbf{u}_2-\mathbf{u}_1$.
If $H_{12}$ is such a reflection, then
$H_{12} J_1(\mathbf{V}_{12}) = \mathbf{V}_{12}$ and
$H_{12} J_2(\mathbf{V}_{21}) = \mathbf{V}_{21}$ which
reduces the problem to the case $\mathbf{A}_{12} = 0$ for $H_{12} J_i$.
\vskip.1cm

\begin{figure}
\begin{center}
\includegraphics[scale=.7]{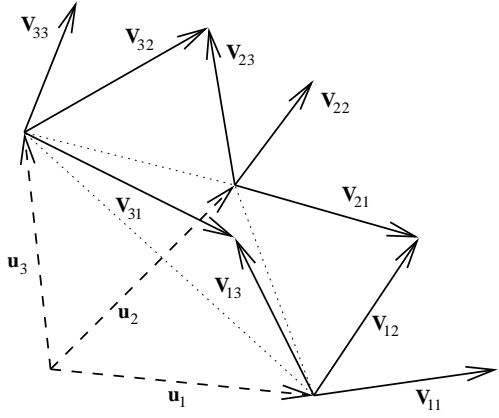}
\caption{Tangent space picture of the nearest neighborhood for the
cubic $\mathbb{Z}^3$ lattice, as determined by a Levi-Civita
connection. The freedom in the choice of such a connection is due
to reflections with respect to hyperplanes through
$\mathbf{u}_i - \mathbf{u}_j$, $i \neq j$.}\label{fig:j3dim}
\end{center}
\end{figure}

Already in three dimensions (see Fig.~\ref{fig:j3dim}) a
classification of the various possibilities turns out
to be quite involved.

\section{Group lattice geometries with torsion}
\label{sec:gl-torsion}
\setcounter{equation}{0}
Section~\ref{sec:discrRiem} demonstrated that Riemannian group
lattices in general do not possess a Levi-Civita connection.
In some cases only \emph{flat} Levi-Civita connections exist
so that one has to allow for non-vanishing torsion
in order to get non-vanishing curvature and thus enough
flexibility to assign a non-trivial geometry to the group
lattice.
\vskip.1cm

Relaxing the previous requirement of vanishing torsion clearly
opens more possibilities for modelling discrete surfaces.
In fact, the following examples demonstrate that linear connections
with torsion naturally appear as properties of Riemannian group
lattice geometries.
The first subsection below shows how in the presence of torsion a
triangle can be curved so that it fits on the surface of a sphere.
The remaining subsections treat some $\mathbb{Z}_4$ lattice examples.

\subsection{A $\mathbb{Z}_3$ lattice geometry with torsion}
\label{subsec:Z3t}
Let $G = \mathbb{Z}_3$ and  $S = \{ 1,2 \}$.
According to section~\ref{sec:tor&curv} the components of the
torsion tensor are given by
\be
 && Q^1{}_{1,1} = 1 + V^1{}_{1,1}  \, , \quad
    Q^1{}_{1,2} = 1 + V^1{}_{1,2}  \, , \quad
    Q^1{}_{2,1} = V^1{}_{2,1}      \, , \quad
    Q^1{}_{2,2} = -1 + V^1{}_{2,2}   \qquad  \nonumber \\
 && Q^2{}_{1,1} = -1 + V^2{}_{1,1} \, , \quad
    Q^2{}_{1,2} = V^2{}_{1,2} \, , \quad
    Q^2{}_{2,1} = 1 + V^2{}_{2,1}  \, , \quad
    Q^2{}_{2,2} = 1 + V^2{}_{2,2}  \, . \qquad
\ee
If we do not require the vanishing of the whole torsion, but only of
the biangle part, i.e. $Q^h_{(0)\, 1,2} = Q^h_{(0)\, 2,1} = 0$, then we
can simulate the geometry of a spherical triangle. Setting
\be
    \g = \left(\begin{array}{cc} 1 & 0 \\ 0 & 1 \end{array} \right)
                \label{Z3_diag_metric}
\ee
a particular solution of the metric-compatibility conditions is
\be
  V_1 = \left(\begin{array}{cc} 0 & -1 \\ 1 & 0 \end{array} \right)
        \, , \qquad
  V_2 = \left(\begin{array}{cc} 0 & 1 \\ -1 & 0 \end{array} \right) \; .
             \label{Z3tor_V}
\ee
Now (\ref{tVell-ell}) leads to
\begin{center}
\begin{tabular}{cc}
\begin{tabular}{c|ccc}
    & at $k+1$ mod $3$ & & at $k$ mod $3$ \\
  \hline
  $\tilde{\V}_{\ell_1} :$ & $\begin{array}{c} \ell_1 \\ \ell_2 \end{array}$ &
  $\mapsto$ & $\begin{array}{r} \ell_2 \\ - \ell_1 \end{array}$
\end{tabular} &
\begin{tabular}{c|ccc}
    & at $k+2$ mod $3$ & & at $k$ mod $3$ \\
  \hline
  $\tilde{\V}_{\ell_2} :$ & $\begin{array}{c} \ell_1 \\ \ell_2 \end{array}$ &
  $\mapsto$ & $\begin{array}{r} - \ell_2 \\ \ell_1 \end{array}$
\end{tabular}
\end{tabular}
\end{center}
which matches the parallel transport along a spherical triangle.
\vskip.1cm

The curvature tensor has only triangle components. Using the matrix
notation $\R_{h_1,h_2} = (\R^h{}_{h',h_1,h_2})$, we obtain
\be
 \R_{(2)\, 1,1} = \left(\begin{array}{cc} -1 & -1 \\
                                           1 & -1
                      \end{array}\right) \, , \qquad
 \R_{(1)\, 2,2} = \left(\begin{array}{cc} -1 & 1 \\
                                          -1 & -1
                      \end{array}\right)
\ee
and $\R_{(0)\, 1,2} = \R_{(0)\, 2,1} = 0$ (vanishing biangle curvature).
The Ricci tensor ${\it Ric}_{h,h'} = \R^1{}_{h,1,h'} + \R^2{}_{h,2,h'}$
is given by
\be
  {\it Ric} = - \left(\begin{array}{cc} 1 & 1 \\
                                      1 & 1
                    \end{array}\right)
\ee
in matrix notation, and the curvature scalar is $R = -2$. The torsion
2-form is given by
\be
  \Theta^1 = \theta^1 \cap \theta^1 \, , \qquad
  \Theta^2 = \theta^2 \cap \theta^2 \; .
\ee
This is an example of a geometry which cannot be isometrically embedded
in a Euclidean space $\mathbb{R}^n$ for any $n \in \mathbb{N}$,
simply due to the fact that with the choice of metric (\ref{Z3_diag_metric})
the sum of the angles of the triangle is $3\pi/2$ and
not $\pi$ as in Euclidean geometry. This fact is taken care of
by the torsion of the connection which causes the backward
parallel transport of the group lattice triangle not to close to
a triangle in the tangent space at a site.
\begin{figure}
\begin{center}
\includegraphics[scale=.6]{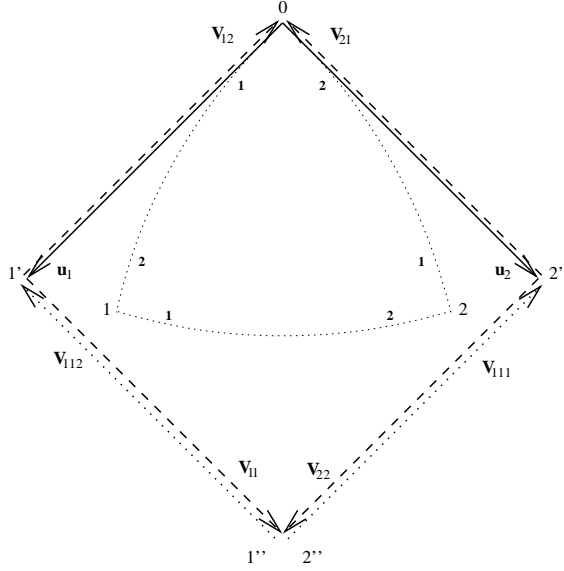}
\caption{The result of backward parallel transport of the group lattice
$(\mathbb{Z}_3, \{ 1,2 \})$ into the tangent space at $0$, using the
connection given by (\ref{Z3tor_V}). The points $1'$ and $1''$,
and also $2'$ and $2''$, do not coincide because of
non-vanishing torsion.}\label{fig:Z3aa}
\end{center}
\end{figure}
The resulting picture in $\mathbb{R}^2$, which represents the tangent
space at the unit element, is drawn in Fig.~\ref{fig:Z3aa}. Here we used
$\mathbf{V}_{11} = -\mathbf{V}_{21} = \mathbf{u}_2$, $\mathbf{V}_{12}
  = -\mathbf{V}_{22} = -\mathbf{u}_1$
which follows from (\ref{Z3tor_V}). The triangle torsion satisfies
\be
     \sum_i \mathbf{u}_i \, Q^i{}_{1,1}
  = \mathbf{u}_1 + \mathbf{V}_{11} - \mathbf{u}_2
  = \mathbf{u}_1 \, , \quad
     \sum_i \mathbf{u}_i \, Q^i{}_{2,2}
  = \mathbf{u}_2 + \mathbf{V}_{22} - \mathbf{u}_1
  = \mathbf{u}_2 \; .
\ee
 From (\ref{V_h_general}) we obtain
\be
 \mathbf{V}_{111} = \mathbf{V}_{221} = - \mathbf{u}_1 \, , \quad
 \mathbf{V}_{112} = \mathbf{V}_{222} = - \mathbf{u}_2
\ee
and, using (\ref{curv&ptransport}) and (\ref{triangle_curv}), the following
curvature expressions:
\be
 \sum_i \mathbf{u}_i \, \R^i{}_{(2)\,1,1,1} = - \mathbf{u}_1 + \mathbf{u}_2
     \, , &&
 \sum_i \mathbf{u}_i \, \R^i{}_{(2)\,2,1,1} = - \mathbf{u}_2 - \mathbf{u}_1
          \nonumber \\
 \sum_i \mathbf{u}_i \, \R^i{}_{(1)\,1,2,2} = - \mathbf{u}_1 - \mathbf{u}_2
     \, ,&&
 \sum_i \mathbf{u}_i \, \R^i{}_{(1)\,2,2,2} = - \mathbf{u}_2 + \mathbf{u}_1
 \, . \qquad
\ee

\subsection{The group lattice $(\mathbb{Z}_4, \{ 1,2 \})$}
\label{subsec:Z412torsion}
\begin{figure}
\begin{center}
\includegraphics[scale=.6]{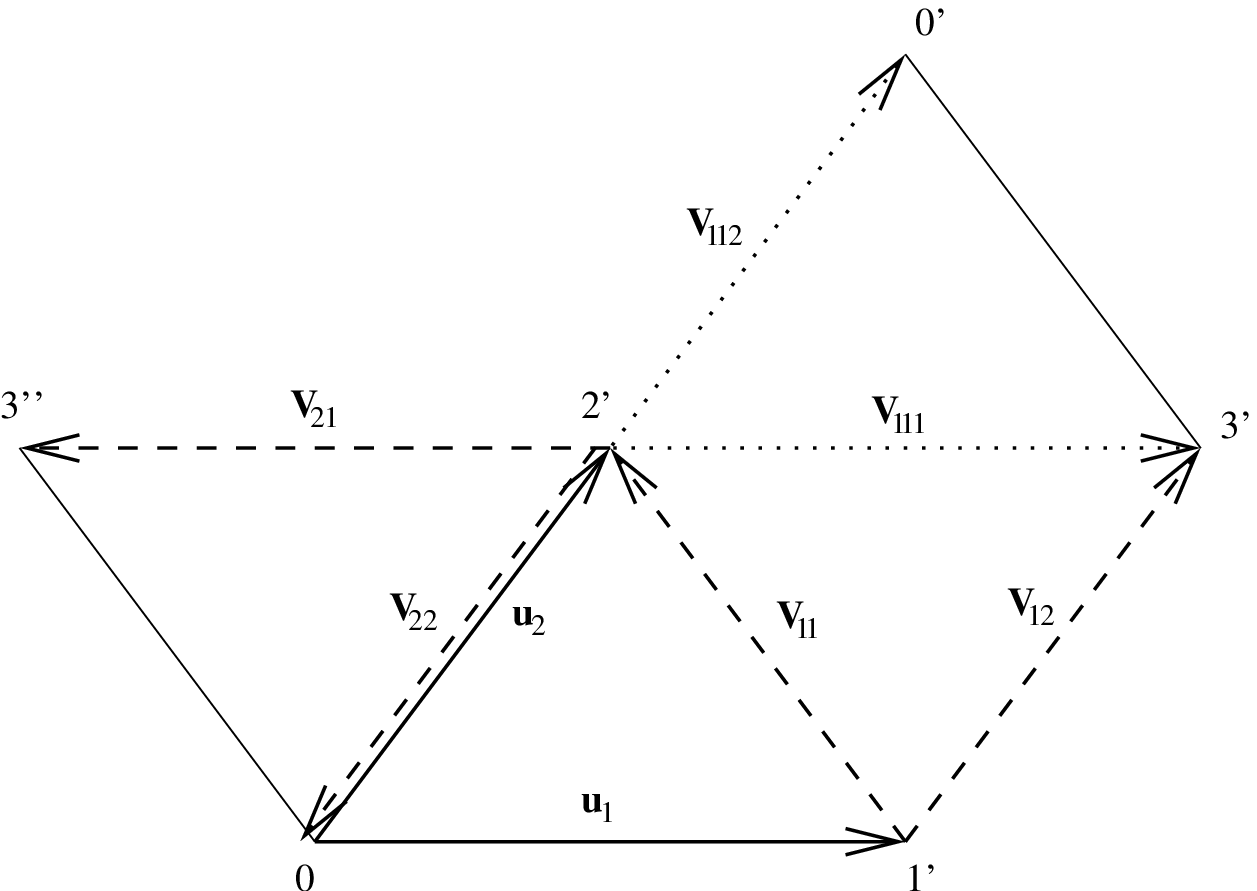}
\caption{The result of backward parallel transport of the group lattice
$(\mathbb{Z}_4, \{ 1,2 \})$ into the tangent space at $0$, using the
connection given by (\ref{V_reg_tetrah_nofold}).}\label{fig:zz412}
\end{center}
\end{figure}

\begin{figure}
\begin{center}
\includegraphics[scale=.6]{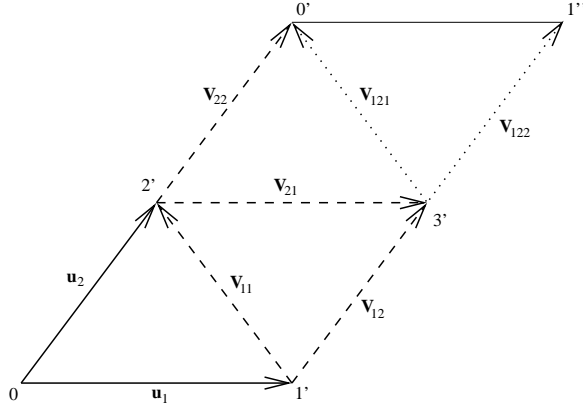}
\caption{The group lattice $(\mathbb{Z}_4, \{ 1,2 \})$ mapped to
the tangent space at $0$ using a connection with non-vanishing
biangle torsion, but vanishing triangle and quadrangle
torsion.}\label{fig:zz412b}
\end{center}
\end{figure}

Let $G = \mathbb{Z}_4$ and $S = \{ 1,2 \}$.
The torsion of a linear connection has the following components:
\be
 \begin{array}{l@{\, = \,}l@{\qquad}l}
 Q^h{}_{2,2} & V^h{}_{2,2} + \dl^h_2  &
                 \mbox{for the biangle 2+2=0}  \\
 Q^h{}_{1,1} & V^h{}_{1,1} - \dl^h_2 + \dl^h_1 &
                 \mbox{for the triangle 1+1=2} \\
 Q^h{}_{2,1} & - Q^h{}_{1,2} = Q^h{}_{1,2;2,1} &  \\
 & V^h{}_{1,2} - V^h{}_{2,1} - \dl^h_2 + \dl^h_1 &
                 \mbox{for the quadrangle 1+2=2+1=3} \, .
 \end{array}
\ee
\vskip.2cm
\noindent
a) If we require vanishing biangle and triangle
torsion, but non-vanishing quadrangle torsion,
the coefficient matrices of the parallel transport have the form
\be
 V_1 = \left(\begin{array}{cc} -1 & p \\ 1 & 1+q \end{array} \right)
       \, , \qquad
 V_2 = \left(\begin{array}{cc} 1+r & 0 \\ s &-1 \end{array} \right)
            \label{z412_Vqtor}
\ee
with functions $p,q,r,s$. As an example, choosing the constant metric
\be
    \g(k)
  = \left( \begin{array}{cc} 1 & 1/2 \\ 1/2 & 1 \end{array} \right)
          \qquad  k=0,1,2,3       \label{g_reg_tetra}
\ee
(which is the metric of a regular tetrahedron surface immersed in
three-dimensional Euclidean space),
and assuming also constant transport matrices, the compatibility
conditions with the connection given by (\ref{z412_Vqtor}) take
the form
\be
   q = p \, , \quad
   p (p+1) = 0 \, , \quad
   s = -1 - r/2 \, , \quad
   r (r+2) = 0
\ee
so that there are four different connections which are compatible
with the metric. All solutions have vanishing biangle curvature.
The solution with $p=-1, r=0$ has non-vanishing triangle
and quadrangle curvature.
The solutions with $(p,r)=(0,0)$, $(p,r)=(-1,-2)$ and
$(p,r)=(0,-2)$ possess only non-vanishing triangle curvature.
In the latter case ($p=0, r=-2$) we have
\be
   V_1 = \left(\begin{array}{cc} -1 & 0 \\
                                  1 & 1
               \end{array}\right)
                                 \, , \qquad
   V_2 = \left(\begin{array}{cc} -1 &  0 \\
                                  0 & -1
               \end{array}\right) \; . \label{V_reg_tetrah_nofold}
\ee
The only non-vanishing part of the curvature 2-form
is the triangle part
\be
   \R_{(2) \, 1,1} = V_1 V_1 - V_2
 = \left(\begin{array}{cc}  2 & 0 \\
                            0 & 2
         \end{array}\right) \; .
\ee
The development of this group lattice in the tangent space at $0$
is shown in Fig.~\ref{fig:zz412}.
Here we used $\mathbf{V}_{11} = \mathbf{u}_2 - \mathbf{u}_1$,
$\mathbf{V}_{12} = \mathbf{u}_2$, $\mathbf{V}_{21} = -\mathbf{u}_1$
and $\mathbf{V}_{22} = -\mathbf{u}_2$ which follows from
(\ref{V_reg_tetrah_nofold}).
The resulting surface does not exhibit folding.
The quadrangle torsion is given by
\be
 \sum_i \mathbf{u}_i \, Q^i{}_{(3)\,1,2} = \mathbf{u}_1 + \mathbf{V}_{12} - \mathbf{u}_2
 - \mathbf{V}_{21} = 2 \, \mathbf{u}_1 \; .
\ee
 Using (\ref{V_h_general}) we obtain
$\mathbf{V}_{111} = \mathbf{u}_1$, $\mathbf{V}_{112} = \mathbf{u}_2$ and thus
the following curvature expressions:
\be
     \sum_i \mathbf{u}_i \, \R^i{}_{(2)\,1,1,1}
  = \mathbf{V}_{111} - \mathbf{V}_{21}
  = 2 \, \mathbf{u}_1  \, , \quad
     \sum_i \mathbf{u}_i \, \R^i{}_{(2)\,2,1,1}
  = \mathbf{V}_{112} - \mathbf{V}_{22}
  = 2 \, \mathbf{u}_2 \; .
\ee
\vskip.1cm
\noindent
{\em Remark.} In general, the compatibility condition for a
constant metric does not enforce a constant connection, i.e.
constant transport matrices. Conversely, a constant connection
may be compatible with non-constant metrics. As an example,
all metrics of the form
\be
 \g(0) = \g(2) =
 \left(\begin{array}{cc} a & b \\ b & c \end{array} \right) \, , \qquad
 \g(1) = \g(3) = \left(\begin{array}{cc} a-2b+c & c-b \\ c-b & c
       \end{array}\right)
\ee
are compatible with the connection (\ref{V_reg_tetrah_nofold}).
\hfill $\blacksquare$
\vskip.2cm
\noindent
b) If only non-vanishing \emph{biangle torsion} is admitted, the coefficient
matrices of the parallel transport take the form
\be
 V_1 = \left(\begin{array}{cc} -1 & p \\ 1 & 1+q \end{array} \right)
       \, , \qquad
 V_2 = \left(\begin{array}{cc} 1+p & u \\ q & v \end{array} \right)
\ee
with functions $p,q,u,v$. If these are taken to be constants, the
compatibility conditions with the metric (\ref{g_reg_tetra}) reduce to
\bez
   \begin{array}{lcl}
    p=q=0    &             &           \\
    u=0, v=1 & \mbox{ or } & u=1, v=-1
   \end{array}
\eez
or
\bez
   \begin{array}{lcl}
    p=q=-1    &             &           \\
    u=1, v=-1 & \mbox{ or } & u=-1, v=0
   \end{array}
\eez
which determines four different connections. The solution
with $(p,q,u,v)=(0,0,0,1)$ has the transport matrices
\be
 V_1 = \left(\begin{array}{cc} -1 & 0 \\ 1 & 1 \end{array} \right)
       \, , \qquad
 V_2 = \left(\begin{array}{cc} 1 & 0 \\ 0 & 1 \end{array} \right)
\ee
for which the curvature 2-form vanishes. The corresponding
tangent space picture obtained by backward parallel transport
of the group lattice into the tangent space at $0$ is drawn
in Fig.~\ref{fig:zz412b}.
Indeed, from the figure we read off $\mathbf{V}_{22} = \mathbf{u}_2$,
$\mathbf{u}_1 + \mathbf{V}_{11} = \mathbf{u}_2$,
$\mathbf{u}_1 + \mathbf{V}_{12} = \mathbf{u}_2 + \mathbf{V}_{21}$,
which determines the above transport matrices.
The biangle torsion satisfies
$\sum_i \mathbf{u}_i \, Q^i{}_{(2)\,1,1} = \mathbf{u}_2 + \mathbf{V}_{22}
 = 2 \, \mathbf{u}_2$.
Furthermore, we have
$\mathbf{V}_{121} = \mathbf{V}_{211} = \mathbf{u}_2 - \mathbf{u}_1$
and $\mathbf{V}_{122} = \mathbf{V}_{212} = \mathbf{u}_2$.
\vskip.1cm

The solution with $(p,q,u,v)=(-1,-1,1,-1)$ has the properties
$V_1 V_1 = V_2$, $V_2 V_2 = V_1$ and $[V_1,V_2]=0$, so that
again the whole curvature 2-form vanishes.
The remaining two solutions have vanishing biangle curvature, but
non-vanishing triangle and quadrangle curvature.
\vskip.2cm
\noindent
c) If only triangle torsion is allowed, there is no connection compatible
with the metric (\ref{g_reg_tetra}).

\subsection{The group lattice $(\mathbb{Z}_4,\{ 1,3 \})$}
\label{subsec:Z413torsion}
\begin{figure}
\begin{center}
\includegraphics[scale=.6]{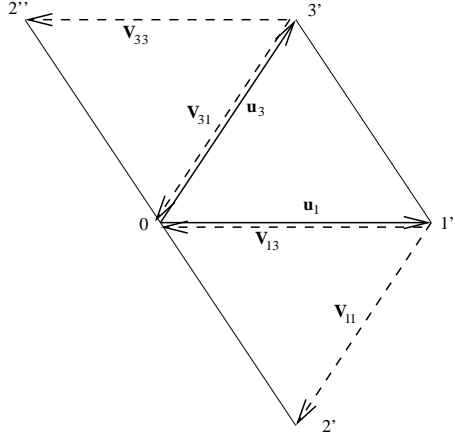}
\caption{The result of backward parallel transport of the group lattice
$(\mathbb{Z}_4, \{ 1,3 \})$ into the tangent space at $0$, using the
connection given by (\ref{z413_Vtelep}).}\label{fig:zz413}
\end{center}
\end{figure}

Let $G = \mathbb{Z}_4$ with $S = \{ 1,3 \}$.
The biangle components of the torsion are
\be
  Q^h{}_{1,3} = \dl^h_1 + V^h{}_{1,3} \, , \qquad
  Q^h{}_{3,1} = \dl^h_3 + V^h{}_{3,1} \, .
\ee
and the quadrangle components (for $g=2$) are
\be
    Q^h{}_{3,3}
  = - Q^h{}_{1,1}
  = Q^h{}_{1,1;3,3}
  = \dl^h_1 - \dl^h_3 + V^h{}_{1,1} - V^h{}_{3,3} \; .
\ee
\vskip.2cm
\noindent
a) Allowing only non-vanishing quadrangle torsion, the parallel
transport matrices have the form
\be
 V_1 = \left(\begin{array}{cc} u & -1 \\ q & 0 \end{array} \right)
       \, , \qquad
 V_3 = \left(\begin{array}{cc} 0 & p \\ -1 & v \end{array} \right)
            \label{z413_Vqtor}
\ee
with functions $p,q,u,v$.
Choosing again the constant metric (\ref{g_reg_tetra})
and assuming constant transport matrices, the compatibility conditions
for the above linear connection reduce to
\be
   q=-1, \; u=0 \quad  \mbox{or} \quad  q = 1, \; u = -1
\ee
and
\be
   p=-1, \; v=0 \quad  \mbox{or} \quad  p = 1, \; v = -1
\ee
which determines four different compatible connections.
The two of them which satisfy $pq = 1$ have vanishing biangle
curvature (\ref{biangle_curv}). One of these, which is given by
\be
   V_1 = V_3 = \left(\begin{array}{cc} 0 & -1 \\
                                      -1 & 0
               \end{array}\right) \, ,
               \label{z413_Vtelep}
\ee
also has vanishing quadrangle curvature (\ref{quadrangle_curv}),
so that the whole curvature 2-form vanishes.
The corresponding development in the tangent space at $0$ is drawn in
Fig.~\ref{fig:zz413} using $\mathbf{V}_{11} = -\mathbf{u}_3$,
$\mathbf{V}_{13} = - \mathbf{u}_1$, $\mathbf{V}_{31} = -\mathbf{u}_3$
and $\mathbf{V}_{33} = -\mathbf{u}_1$.
There is no folding.
The quadrangle torsion is given by
$\sum_i \mathbf{u}_i \, Q^i{}_{(2)\,1,1}
  = \mathbf{u}_1 + \mathbf{V}_{11} - \mathbf{u}_3 - \mathbf{V}_{33}
  = 2 \, (\mathbf{u}_1 - \mathbf{u}_3)$.
\vskip.2cm
\noindent
b) If we require vanishing quadrangle torsion, but allow for
non-vanishing \emph{biangle torsion}, the parallel transport matrices
take the form
\be
 V_1 = \left(\begin{array}{cc} u & q \\ 1+v & p \end{array} \right)
       \, , \qquad
 V_3 = \left(\begin{array}{cc} r & 1+u \\ s & v \end{array} \right)
            \label{z413_Vbtor}
\ee
with functions $p,q,r,s,u,v$. If these are constants, the compatibility
conditions with the constant metric (\ref{g_reg_tetra}) have the
following solutions,
\bez
   \begin{array}{lcl}
    u=v=0    &             &           \\
    p=0, q=1 & \mbox{ or } & p=1, q=-1 \\
    r=0, s=1 & \mbox{ or } & r=1, s=-1
   \end{array}
\eez
and
\bez
   \begin{array}{lcl}
    u=v=-1    &             &           \\
    p=1, q=-1 & \mbox{ or } & p=-1, q=0 \\
    r=1, s=-1 & \mbox{ or } & r=-1, s=0
   \end{array}
\eez
which determine eight different connections. Among them there are
three solutions with vanishing biangle curvature:
\bez
   (p,q,r,s,u,v) \in
   \{ (0,1,0,1,0,0), \, (1,-1,1,-1,0,0), \, (-1,0,-1,0,-1,-1) \} \; .
\eez
For the first solution we obtain
\be
 V_1 = V_3 = \left(\begin{array}{cc} 0 & 1 \\ 1 & 0 \end{array} \right)
             \label{z413_Vbtor_reg}
\ee
and for the third
\be
   V_1 = V_3
 = \left(\begin{array}{cc} -1 & 0 \\ 0 & -1 \end{array} \right) \; .
\ee
For both also the quadrangle curvature and thus the whole
curvature 2-form vanishes.
The second solution has more complicated transport matrices:
\be
 V_1 = \left(\begin{array}{cc} 0 & -1 \\ 1 & 1 \end{array} \right)
       \, , \qquad
 V_3 = \left(\begin{array}{cc} 1 & 1 \\ -1 & 0 \end{array} \right) \, .
\ee
The corresponding quadrangle curvature is
\be
   \R_{(2)\,3,3} = - \R_{(2) \, 1,1} = \R_{(2) \, 1,1;3,3}
 = \left(\begin{array}{cc}
   -1 & -2 \\ 2 & 1 \end{array} \right) \; .
\ee
The no-folding conditions are $\det V_1 < 0$ and $\det V_3 < 0$.
They select the transport matrices (\ref{z413_Vbtor_reg}).

\section{Group lattice geometry and coordinates}
\label{sec:coord}
\setcounter{equation}{0}
In order to explore discrete structures in close analogy with the continuum
it should be of some interest to consider analogs of coordinates
and coordinate transformations, as well as the associated properties of
geometric objects. Moreover, if there is a continuum limit,
as in the case of a hypercubic $\mathbb{Z}^n$ lattice,
one should recover the corresponding continuum structures.
\vskip.1cm

Let $(G,S)$ be a group lattice with $|S| = n$. Real functions $x^\mu$,
$\mu=1,\ldots,n$, are said to be \emph{coordinates} on $G$ if
$(x^\mu) : G \to \mathbb{R}^n$ is injective and the matrix $(\ell_h x^\mu)$ is
invertible at all $g \in G$. If coordinates do not exist globally,
they can still be introduced on subsets of $G$.
\vskip.1cm

The first subsection below presents an example of a coordinate system on
a $\mathbb{Z}_4$ lattice. In particular, it demonstrates a relation
between discrete structures and noncommutative differential calculi
on the algebra of functions on $\mathbb{R}^n$ which has not yet been
sufficiently explored.
The second subsection then treats in some detail Riemannian geometry of
a hypercubic $\mathbb{Z}^n$ lattice in terms of adapted coordinates.

\subsection{Coordinates on $(\mathbb{Z}_4, \{ 1,2 \})$}
The two functions
\be
   x = e^0 - e^1 + e^2 - e^3 \, , \qquad
   y = e^0 + e^1 - e^2 - e^3
\ee
are coordinates on $\mathbb{Z}_4$ with $S = \{ 1,2 \}$. Since
$(x(0),y(0)) = (1,1)$, $(x(1),y(1)) = (-1,1)$, $(x(2),y(2)) = (1,-1)$
and $(x(3),y(3)) = (-1,-1)$, the map
$(x,y) : \mathbb{Z}_4 \rightarrow \mathbb{R}^2$ is obviously injective.
Using
\be
  R^\ast_1 x = -x \, , \quad R^\ast_2 x = x \, , \quad
  R^\ast_1 y = x \, y \, , \quad R^\ast_2 y = -y
\ee
we obtain the Jacobian
\be
    (\ell_h x^\mu)
  = \left( \begin{array}{cc} -2 \, x & 0 \\
                         (x-1) \, y & -2 \, y
           \end{array} \right)
\ee
which is indeed invertible at each lattice site. Every function on $\mathbb{Z}_4$
can be expressed as a function of $x$ and $y$. They satisfy
\be
      x^2 = y^2 = \mathbf{1}  \; .  \label{x2=1=y^2}
\ee
The coordinates $x,y$ then constitute a representation of $\mathbb{Z}_4$.
For the differentials we obtain the expressions
\be
  \d x = [\theta,x] = - 2 \, x \, \theta^1 \, , \quad
  \d y = [\theta,y] = (x-1) \, y \, \theta^1 - 2 \, y \, \theta^2
\ee
and thus, using $x^2 = \mathbf{1}$,
\be
  \theta^1 = - {1 \over 2 \, x} \, \d x \, , \quad
  \theta^2 = {1 \over 4} \, (x-1) \, \d x - {1 \over 2 \, y} \, \d y \; .
\ee
Furthermore, using $\theta^h f = R_h^\ast f \, \theta^h$ we obtain the
following commutation relations between the coordinates $x,y$ and
their differentials:
\be
   [\d x,x] = -2 \, x \, \d x \, , \quad
   [\d y,y] = - 2 \, y \, \d y \, , \quad
   [\d x,y] = [\d y,x] = (x-1) \, y \, \d x \; .
     \label{coord_dc_Z4_12}
\ee
We have thus reached a formulation of the differential calculus on
$(\mathbb{Z}_4, \{ 1,2 \})$ as a noncommutative differential calculus
on $\mathbb{R}^2$. Indeed, imposing the relations (\ref{coord_dc_Z4_12})
on two real functions $x,y$, the group lattice $(\mathbb{Z}_4, \{ 1,2 \})$
can be essentially recovered. The first two relations imply
$\d(x^2) = 0 = \d(y^2)$. As a consequence, $x^2$ and $y^2$ are ``constants''
for this differential calculus and commute with differentials.
Using (\ref{coord_dc_Z4_12}) this implies
\be
 0 &=& [\d(y^2),x] = [\d x,y^2] = [\d x,y] \, y + y \, [\d x,y]
    = (x-1) \, y \, (\d x) \, y + (x-1) \, y^2 \, \d x \nonumber \\
   &=& (x-1) \, y \, xy \, \d x + (x-1) \, y^2 \, \d x
    = (x^2-1) \, y^2 \, \d x
\ee
and thus $x^2 = \mathbf{1}$, assuming $y^2 \neq 0$ and that
$\Omega^1$ is free with basis $\d x, \d y$.
The equations (\ref{coord_dc_Z4_12}) are homogeneous in $y$,
so that they are not able to fix the value of $y^2$.
But the calculus is obviously consistent with the constraint $y^2 = \mathbf{1}$.
Passing over to the algebra $\A$ of functions generated by the
variables $x,y$ modulo the relations (\ref{x2=1=y^2}) and setting
\be
  e^0 = {(1+x)(1+y) \over 4} \, , \; e^1 = {(1-x)(1+y) \over 4} \, , \;
  e^2 = {(1+x)(1-y) \over 4} \, , \; e^3 = {(1-x)(1-y) \over 4}  \quad
\ee
we find $e^i e^j = \dl^{i,j} \, e^i$ and $\sum_i e^i = 1$. These are
the primitive idempotents of $\A$.
\vskip.1cm

Let us deduce some more consequences from the commutation relations
(\ref{coord_dc_Z4_12}). They are equivalent to
\be
  \d x \; x = - x \, \d x \, , \quad
  \d x \; y = x \, y \, \d x \, , \quad
  \d y \; y = - y \, \d y \, , \quad
  \d y \; x = x \, \d y + (x-1) \, y \, \d x
\ee
so that
\be
  \d x \; f(x,y) &=& f(-x,xy) \, \d x  \\
  \d y \; f(x,y) &=& f(x,-y) \, \d y + {f(x,xy)-f(-x,xy) \over 2 \, x}
                   \, (x-1) \, y \, \d x \, . \label{coord_dy_f}
\ee
Introducing (left) partial derivatives of a function $f$ via
\be
    \d f = \pa_x f \; \d x + \pa_y f \; \d y
\ee
we find
\be
   \d y \; f(x,y) - f(x,y) \, \d y = [\d f,y]
   = (\pa_x f) \, [\d x,y] + (\pa_y f) \, [\d y,y]
\ee
which together with (\ref{coord_dc_Z4_12}) and (\ref{coord_dy_f}) leads to
\be
  \pa_x f = {1 \over 2x} \, \Big( f(x,xy) - f(-x,xy) \Big) \, , \quad
  \pa_y f = {1 \over 2y} \, \Big( f(x,y) - f(x,-y) \Big) \; .
\ee
A similar calculation starting with $\d x \, f(x,y) - f(x,y) \, \d x
= [\d f,x]$ leads to an apparently different expression for $\pa_x f$.
It reduces to the above formula with the help of
\be
   f(x,xy) = {1 \over 2} \Big( (x+1) \, f(x,y) - (x-1) \, f(x,-y) \Big)
\ee
which holds as a consequence of $x^2 = \mathbf{1}$.
\vskip.1cm

Of course, all geometric structures on $(\mathbb{Z}_4, \{ 1,2 \})$
can now be expressed in terms of the coordinates and their differentials.

\subsection{Hypercubic group lattice geometry in coordinates}
\label{subsec:Z^n}
Let $G$ be the additive group $\mathbb{Z}^n$ and $S = \{ \hmu | \,
\mu = 1,\ldots,n \}$ the standard basis of $\mathbb{Z}^n$,
i.e., $\hmu = (0,\ldots,0,1,0,\ldots,0)^T$ with the $1$ at the
$\mu$th position.
There are no biangles or triangles, but only quadrangles.
The group lattice is the oriented hypercubic lattice and for $a \in \mathbb{Z}^n$
the functions $e^a$ form a basis over $\mathbb{C}$ of $\A$. Then
$(\ell_\hmu f)(a) = f(a+\hmu)-f(a)$ defines a basis $\{ \ell_\hmu \}$
of the space $\X$ of vector fields. The dual basis of $\Omega^1$ is given by
\be
 \theta^\hmu = \sum_{a \in \mathbb{Z}^n} e^a \, \d e^{a+\hmu} \, .
\ee
The functions $x^\mu = \kappa \, \sum_{a \in \mathbb{Z}^n} a^\mu e^a$,
$\mu = 1,\ldots,n$, with a constant $\kappa$, are coordinates on the space.
Every function can be written as $f(x)$ with $x = (x^1,\ldots,x^n)$.
Furthermore, we find
\be
  \theta^\hmu = {1 \over \kappa} \, \d x^\mu  \qquad \mu = 1,\ldots,n \, .
\ee
Since $\hmu+\hnu = \hnu+\hmu$, the 2-form relations
$\d x^\mu \, \d x^\nu = - \d x^\nu \, \d x^\mu$
hold for all pairs $\mu,\nu = 1,\ldots,n$. As a consequence, every product
of the form $\d x^{\mu_1} \cdots \d x^{\mu_r}$ is totally antisymmetric.
Since the group is Abelian,
$\d x^{\mu_1} \cap \cdots \cap \d x^{\mu_r} = \d x^{\mu_1} \cdots \d x^{\mu_r}$.
This implies that $\alpha_1 \cap \cdots \cap \alpha_r$ is totally antisymmetric
for arbitrary 1-forms $\alpha_i$. It should be noticed, however, that
$\alpha_1 \cdots \alpha_r$ is not antisymmetric, in general.
\vskip.1cm

Introducing (left) partial derivatives of a function $f(x)$ via
\be
     \d f = \sum_{\mu=1}^n (\pa_{+\mu} f) \, \d x^\mu
\ee
we find
\be
  \pa_{+\mu}f = { R^\ast_\hmu f - f \over \kappa} \, , \qquad
  (R^\ast_\hmu f)(x) = f(x + \kappa \hmu) \; .
\ee
The backward parallel transport of a linear connection with
transport matrices $V_\mu = (V^\rho{}_{\mu\sigma})$ acts as follows,
\be
   \tilde{\V}_{\pa_{+\mu}} \pa_{+\nu}
 = {1 \over \kappa} \sum_\rho V^\rho{}_{\mu\nu} \cdot \pa_{+\rho} \; .
    \label{Zn_tVpa}
\ee
Let us write
\be
      \na \d x^\mu
  = \theta \oA \d x^\mu - \sum_{\nu=1}^n V^\mu{}_\nu \oA \d x^\nu
  = - \sum_{\nu = 1}^n \Gamma^\mu{}_\nu \oA \d x^\nu
\ee
where
\be
    V^\mu{}_\nu
 = {1 \over \kappa} \sum_{\rho=1}^n V^\mu{}_{\rho\nu} \, \d x^\rho
     \, , \quad
   \Gamma^\mu{}_\nu
 = \sum_{\rho=1}^n \Gamma^\mu{}_{\rho \nu} \, \d x^\rho  \; .
\ee
Using
$\theta = \sum_{\mu=1}^n \theta^\hmu = (1/\kappa)
\sum_{\mu=1}^n \vartheta_\mu \, \d x^\mu$
with $\vartheta_\mu = 1$ for $\mu=1,\ldots,n$, we obtain
\be
     \Gamma^\mu{}_{\rho\nu} = {1 \over \kappa} \,
       [ V^\mu{}_{\rho\nu} - \vartheta_\rho \, \dl^\mu_\nu ]  \; .
\ee
For a suitable Levi-Civita connection these functions should yield the
Christoffel symbols in the continuum limit.
\vskip.1cm

The components of the torsion 2-form
$\Theta^\mu = (1/2) \sum_{\nu,\rho=1}^n Q^\mu{}_{\nu \rho} \,
\d x^\nu \cap \d x^\rho$ are
\be
  Q^\mu{}_{\nu\rho} = {1 \over\kappa} \, (V^\mu{}_{[\nu\rho]} -
                      \vartheta_{[\nu} \, \dl^\mu_{\rho]})
                    = \Gamma^\mu{}_{[\nu\rho]}
\ee
and the components of the curvature 2-form
$\R(\d x^\mu) = (1/2) \sum_{\nu,\rho,\sigma=1}^n
\R^\mu{}_{\nu\rho\sigma} \, \d x^\rho \cap \d x^\sigma \oL \d x^\nu$
are given by
\be
    \R^\mu{}_{\nu \rho \sigma} = (\R_{\rho \sigma})^\mu{}_\nu
  = {1 \over\kappa^2} \, ( V_\rho \: R^\ast_\hrho V_\sigma-V_\sigma \:
    R^\ast_{\hat{\sigma}} V_\rho )^\mu{}_\nu \; .
\ee
\vskip.1cm

The two Bianchi identities take the form
\be
  {1 \over \kappa} \left(\sum_{\lambda=1}^n
   V^\mu{}_{[\nu|\lambda|} \; R^\ast_{\hnu}Q^\lambda{}_{\rho\sigma]}
   - Q^\mu{}_{[\nu\rho} \; R^\ast_{\hnu+\hrho}\vartheta_{\sigma]} \right)
   = \R^\mu{}_{[\nu\rho\sigma]}
\ee
and
\be
    V_{[\nu}\; R^\ast_\hnu\R_{\rho\sigma]}
  = \R_{[\nu\rho}\; R^\ast_{\hnu+\hrho}V_{\sigma]} \, .
\ee
\vskip.1cm

The compatibility condition for the linear connection and a metric tensor
\be
  \mathsf{g} = \sum_{\mu,\nu=1}^n \g_{\mu\nu}(x) \, \d x^\mu \oL \d x^\nu
\ee
reads
\be
   R^\ast_\hrho \, (\g_{\mu\nu}) = V^T_\rho \, (\g_{\mu\nu}) \, V_\rho  \; .
\ee
The integrability condition of this equation (iteration around a plaquette)
implies that the matrices $K_{\mu\nu}$ which are defined by
\be
  V_\mu \; R^\ast_\hmu V_\nu = K_{\mu\nu} \, V_\nu \; R^\ast_\hnu V_\mu
\ee
are isometries of $\g$ at every point of the lattice.
The curvature tensor, in matrix form, can now be written as follows,
\be
   \R_{\mu\nu}
 = {1 \over \kappa^2} \, (K_{\mu\nu}-I) \, V_\nu \; R^\ast_\hnu V_\mu \, .
\ee
If the torsion vanishes, the first Bianchi identity reduces to
$\R^\mu{}_{[\nu\rho\sigma]} = 0$. Then there is (up to the global sign)
only one definition of a Ricci tensor:
\be
  {\it Ric}_{\mu\nu} = \sum_{\rho=1}^n \R^\rho{}_{\mu\rho\nu} \, .
\ee
The curvature scalar is given by
\be
 \R = \sum_{\mu,\nu=1}^n \g^{\mu\nu} \, {\it Ric}_{\mu\nu}
\ee
involving the components of the inverse metric
$\mathsf{g}^{-1} = \sum_{\mu,\nu=1}^n \g^{\mu\nu} \, \pa_{+\mu} \oL \pa_{+\nu}$.
\vskip.2cm

Let $y^\mu(x)$ be a set of $n$ real-valued functions which can be
inverted to express $x^\nu$ in terms of the functions $y^\mu$ and
for which the Jacobian
\be
       {\cal J}^\mu{}_\nu := \pa_{+\nu} y^\mu
\ee
is invertible. The functions $y^\mu$ are then new coordinates
and we have
\be
   \d y^\mu = \sum_{\nu = 1}^n {\cal J}^\mu{}_\nu \, \d x^\mu \, , \qquad
   \d x^\mu = \sum_{\nu = 1}^n ({\cal J}^{-1})^\mu{}_\nu \, \d y^\nu \; .
\ee
Note that
$\d y^\mu \cap \d y^\nu + \d y^\nu \cap \d y^\mu = 0$, while
$\d y^\mu \, \d y^\nu + \d y^\nu \, \d y^\mu \neq 0$, in general.
Introducing (left) partial derivatives with respect to the
basis $\d y^\mu$ via
\be
  \d f = \sum_{\nu=1}^n \pa^y_{+\nu} f \, \d y^\nu
\ee
we obtain
\be
  \pa^y_{+\nu}f = \sum_{\mu=1}^n({\cal J}^{-1})^\mu{}_\nu \, \pa_{+\mu}f
\ee
and, in particular,
\be
   ({\cal J}^{-1})^\mu{}_\nu = \pa^y_{+\nu} x^\mu  \; .
\ee
Using the coordinates $x^\mu$, the basic commutation relations
of the differential calculus are
\be
  [\d x^\mu,x^\nu] = \kappa \, \dl^{\mu\nu} \, \d x^\mu \; .
\ee
In terms of $y^\mu$ they read
\be
  [\d y^\mu,y^\nu] = \kappa \, \sum_{\rho=1}^n C^{\mu\nu}{}_\rho \, \d y^\rho \, ,
  \qquad
  C^{\mu\nu}{}_\rho := \sum_{\sigma=1}^n {\cal J}^\mu{}_\sigma \,
  {\cal J}^\nu{}_\sigma \, ({\cal J}^{-1})^\sigma{}_\rho \; .
\ee
In the limit as $\kappa \to 0$ we obtain in both coordinate systems the ordinary
continuum differential calculus, as long as the coordinate transformation does
not involve $\kappa$. If $f$ and $y^\mu$ are differentiable functions
of $x^\mu$, then in this limit $\d f$ becomes $\sum_\mu(\pa f/\pa x^\mu) \, \d x^\mu$
and also $\sum_\mu (\pa f/\pa y^\mu) \, \d y^\mu$ with the help of the chain rule.
Although the lattice differential calculus becomes particularly simple when expressed
in terms of the coordinates $x^\mu$, in the continuum limit all coordinate
systems are on an equal footing.
The discrete calculus also allows $\kappa$-dependent coordinate transformations.
But exploring the continuum limit we should require that such a transformation
remains a coordinate transformation in the limit $\kappa \to 0$.
\vskip.1cm

Since the metric is defined using the left-covariant tensor product, the metric
components transform homogeneously with the Jacobi matrix:
\be
   \g'_{\mu\nu}(y)
 = \sum_{\rho,\sigma=1}^n ({\cal J}^{-1})^\rho{}_\mu \,
   ({\cal J}^{-1})^\sigma{}_\nu \, \g_{\rho\sigma}(x)
\ee
where $\mathsf{g} = \sum_{\mu,\nu} \g'_{\mu\nu}(y) \, \d y^\mu \oL \d y^\nu$.
This local tensor transformation property is shared by the components of
the torsion and curvature, in particular. A linear connection and the
associated transport matrices have a non-local character.
With the help of (\ref{tVell-ell}) and (\ref{Zn_tVpa}) we find
\be
  V'_\mu(y) = \sum_{\nu=1}^n ({\cal J}^{-1})^\nu{}_\mu(x) \, {\cal J}(x) \,
              V_\nu(x) \, {\cal J}^{-1}(x + \kappa \hnu) \; .
\ee

\section{Conclusions}
\label{sec:concl}
\setcounter{equation}{0}
Starting from basic formulas of noncommutative geometry, we
developed a formalism of Riemannian geometry
of group lattices. More precisely, we restricted our
considerations to the subclass of bicovariant group lattices.
Only for this subclass there is a simple conversion between the
ordinary tensor product $\oA$ and the left-covariant tensor
product $\oL$. The latter played a crucial role in making
contact with classical geometry. In particular, it allows
to introduce a discrete analogue of a metric tensor with
a natural geometric interpretation and, more technically,
to formulate a compatibility condition with a linear connection.
\vskip.1cm

In particular in the case of a $\mathbb{Z}^n$ group lattice,
the discrete geometry obtained has much in common with lattice
gauge theory. It yields a discretization of continuum geometry
via plaquettes where the curvature results from parallel transport
around a plaquette (see also the various approaches \cite{latt_grav}
to ``lattice gravity'' in this context). In contrast, in Regge
calculus the curvature is concentrated at a hinge (which in two
dimensions is a vertex).
\vskip.1cm

Given a metric, the compatibility condition for a linear
connection leaves us with the freedom of torsion. This is
analogous to continuum differential geometry where the
additional requirement of vanishing torsion uniquely
determines a particular linear connection, the Levi-Civita
connection (which is expressed via the Christoffel symbols
in terms of the metric coefficients). The situation is much
more complicated for group lattices, however.
\vskip.1cm

A Levi-Civita connection need not exist for a given Riemannian
group lattice. Furthermore, if such a connection exists, then
it is not unique. We achieved a geometric understanding
of this ambiguity through the elaboration of several examples.
The deeper origin is the fact that our connections have values
in a group algebra rather than a Lie algebra. The latter only
feels the part of a (continuous) group which is connected
with the identity. The requirement of a continuum limit
in general distinguishes a certain connection.
\vskip.1cm

The requirement of a Levi-Civita connection for a Riemannian
group lattice strongly restricts the metric, in general.
On the other hand, we learned from our examples that
metric-compatible linear connections with non-vanishing torsion
show up quite naturally. A convenient condition which replaces
that of vanishing torsion is not available.
A few general statements can nevertheless be made.
Vanishing triangle torsion means assigning Euclidean
properties to the respective triangle. Of course, a group lattice
with Euclidean triangles, but more than three sites, may still be curved.
Non-vanishing biangle torsion allows for an anisotropy of the
distance relation between the respective two lattice sites,
adhering to a simple interpretation of the metric coefficients.
The requirement of vanishing biangle torsion would rule out
this feature. But it would also eliminate geometries without
such an anisotropy as we saw in sections~\ref{subsec:Z412torsion}
and \ref{subsec:Z413torsion}.
\vskip.1cm

On the other hand,
a distance anisotropy may indeed appear in communication networks
(with a group lattice structure), a relation which should be
elaborated elsewhere. \cite{cnetwork}
The design of a communication network determines
its efficiency. The broadcast time, for example, clearly depends
on its geometry. \cite{DFV91}
For such problems the geometric formalism developed in this
work could be of help.
\vskip.1cm

Our examples demonstrate that torsion quite naturally enters the stage.
The more we depart from the continuum, the more we get away from the
familiar condition of vanishing torsion of continuum (pseudo-)
Riemannian geometry. Hypercubic group lattices, which only consists
of quadrangles, are relatively close to the continuum in this sense.
Biangles and triangles add to the rigidity of a lattice, so that
torsion becomes more necessary in order to curve it.
The conclusion is that, in contrast to ordinary
continuum differential geometry, (non-zero) torsion is an essential
ingredient of our discrete geometric formalism. Interesting field
equations will have to take care of this fact and describe the dynamics
of metric \emph{and} torsion.
\vskip.1cm

Is there a distinguished geometry associated with a (bicovariant)
group lattice? Indeed, a direct consequence of the definition of
a group lattice is the existence of a family of vector fields
$\ell_h$, $h \in S$. Requiring that these are Killing vector fields
of the metric, so that their flows preserve the metric, restricts
the a priori possible metrics to the class of right-invariant
metrics which are completely determined by the components at
one site. If $S$ is Abelian, these are simply the constant metrics,
i.e., the components are the same at all sites (which correspond to
the group elements).
Associated with the class of right-invariant metrics is a distinguished
metric-compatible linear connection. Moreover, we have the notion
of bi-invariance of a metric which determines a subclass of
right-invariant metrics.
Interesting relations between group structure and geometry are
expected to emerge from this.
\vskip.1cm

Even in the familiar hypercubic lattice case the (pseudo-) Riemannian
geometry derived from the general framework of group lattice geometry
appears to be new. In particular in the form presented in
section~\ref{subsec:Z^n}, using coordinates on the lattice,
the close analogy with continuum Riemannian geometry becomes
transparent. This provides an alternative to the existing
discretizations of gravity theories.
\vskip.1cm

Representations of ``intrinsic'' group lattice geometries
via immersions in a Euclidean space will be treated in a separate work.
For two-dimensional Riemannian group lattices (where
$S$ consists of two different elements), the bicovariance condition
restricts to Abelian groups, and a relatively simple formalism of
immersions can be developed in analogy with that of continuum
differential geometry. For immersions of higher than two-dimensional
Riemannian group lattices in Euclidean $\mathbb{R}^n$ the formalism
is more complex and new features will show up.

\begin{appendix}
\renewcommand{\theequation} {\Alph{section}.\arabic{equation}}
\addcontentsline{toc}{section}{\numberline{}Appendices}

\section{Orthonormal coframe fields}
\label{sec:ocf}
\setcounter{equation}{0}
Let $\mathsf{g}$ be a metric on a group lattice $(G,S)$ which has Euclidean
(or Lorentzian) signature at each point. An {\em orthonormal coframe field}
is a set of $|S|$ linearly independent 1-forms $E^a$ (at each point of $G$)
such that
\be
   \mathsf{g} = \sum_{a,b=1}^{|S|} \eta_{ab} \, E^a \oL E^b
\ee
where $\eta = (\eta_{ab})$ has entries $\pm 1$ on the diagonal and zeros
otherwise (according to the signature of $\mathsf{g}$). Writing
\be
    E^a = \sum_{h \in S} E^a{}_h \, \theta^h \, , \qquad a = 1,\ldots,|S|
\ee
it follows that the matrix $(E^a{}_h)$ is invertible at all sites $g \in G$.
Let $(\bar{E}^h{}_a)$ denote its inverse.
In the following, for $(\eta_{ab})$ we may take more generally an
arbitrary constant symmetric matrix.
Using (I.6.5) and (\ref{Vellh_thetah}), we find
\be
 \V_{\ell_h} E^a = \sum_{b=1}^{|S|} (R^\ast_{h^{-1}} L^a{}_{h,b}) \, E^b
        \label{VellhE^a}
\ee
with
\be
   L^a{}_{h,b} := \sum_{h',h'' \in S} E^a{}_{h'} \, V^{h'}{}_{h,h''} \,
                 R_h^\ast \bar{E}^{h''}{}_b \; .
\ee
or $L_h = E \, V_h \, R^\ast_h \bar{E}$ in an obvious matrix notation.
As a consequence,
\be
   \nabla E^a = \theta \oA E^a - \sum_b L^a{}_b \oA E^b
\ee
with
\be
    L^a{}_b := \sum_{h \in S} L^a{}_{h,b} \, \theta^h \; .
\ee
\vskip.2cm

Let us introduce the dual frame field
\be
   \bar{E}_a := \sum_{h \in S} \bar{E}^h{}_a \cdot\ell_h
\ee
which satisfies $\langle \bar{E}_a , E^b \rangle = \delta_a^b$.
As a consequence of (I.7.17) and (\ref{tVell-ell}), we find
\be
   \tilde{\V}_{\ell_h} \bar{E}_a = \sum_b L^b{}_{h,a} \cdot \bar{E}_b
   \; .
\ee
\vskip.1cm

The metric-compatibility condition for the connection takes the form
\be
    L^T_h \, \eta \, L_h = \eta  \; .
\ee
The matrices $L_h$ are thus isometries of $\eta$, they have values
in the orthogonal group $O(\eta)$ of $\eta$. This shows that if an
orthonormal coframe field is chosen, \emph{an $\eta$-compatible linear
connection is equivalent to a map $G \times S \rightarrow O(\eta)$}.
\vskip.1cm

The components of the torsion 2-form with respect to the coframe $E^a$ are
\be
   \Theta^a := \Theta(E^a)
 = E^a \, \theta - \Delta(E^a) + \sum_{b=1}^{|S|} L^a{}_b \, E^b  \; .
\ee
Here we used (I.7.6), $\Delta( f \, \omega ) = f \, \Delta(\omega)$,
(I.6.5) and (\ref{VellhE^a}). Writing this as
\be
  \Theta^a = \sum_{h_1,h_2 \in S} \Big( E^a{}_{h_1}
             - \sum_{h \in S} E^a{}_h \, \delta^h_{h_1h_2}
             + \sum_b L^a{}_{h_1,b} \, R_{h_1}^\ast E^b{}_{h_2} \Big)
             \, \theta^{h_1} \, \theta^{h_2}
\ee
the condition of vanishing torsion $\Theta^a = 0$ yields for biangles
($h_1 h_2=e$, $h_1,h_2 \in S_{(0)}$)
\be
  L_{h_1} \, R^\ast_{h_1} E_{h_2} = - E_{h_1}
\ee
where, for example, $E_{h_1}$ denotes the column with entries $E^a{}_{h_1}$.
For triangles ($h_1h_2 = h \in S_{(1)}$) it yields
\be
  L_{h_1} \, R^\ast_{h_1} E_{h_2} = E_h - E_{h_1}
\ee
and for quadrangles ($h_1 h_2 = \hat{h}_1 \hat{h}_2 = g \in S_{(2)}$)
\be
  L_{h_1} \, R^\ast_{h_1} E_{h_2} - L_{\hat{h}_1} \, R^\ast_{\hat{h}_1} E_{\hat{h}_2}
  = E_{\hat{h}_1} - E_{h_1} \, .
\ee
\vskip.1cm

The components of the curvature with respect to the coframe $E^a$ are
\be
    \R(E^a) = \sum_{b=1}^{|S|} \R^a{}_b \oA E^b
\ee
where
\be
    (\R^a{}_b) = L^2 - \Delta(L) - I \, \Delta^e    \, , \qquad
    L := \sum_{h \in S} L_h \, \theta^h   \; .
\ee
With the help of the Leibniz rule and (I.2.15), we obtain the first
Bianchi identity (I.7.11) in the following form,
\be
     \d \Theta(E) + (L - \theta) \, \Theta(E)
   = \R^a{}_b \, E^b
   = - \Delta^e E - \Delta(L) \, E + L^2 \, E
\ee
where $E$ stands for the column with entries $E^a$. From
\be
 0 = \nabla(\R(E^a)) - \R(\nabla E^a)
   = \sum_b \Big( - \Delta(\R^a{}_b) + \sum_c ( L^a{}_c \, \R^c{}_b
     - \R^a{}_c \, L^c{}_b ) \Big) \oA E^b
\ee
we obtain the following version of the second Bianchi identity,
\be
    \Delta( \R^a{}_b )
  = \sum_c \Big( L^a{}_c \, \R^c{}_b - \R^a{}_c \, L^c{}_b \Big) \; .
\ee
Writing
\be
  \R(E^a) = \sum_{b=1}^{|S|} \sum_{h_1,h_2 \in S} \R^a{}_{b,h_1,h_2} \,
            \theta^{h_1} \cap \theta^{h_2} \oL E^b
\ee
we find the biangle part of the curvature
\be
   \R_{(e) \, h_1,h_2}
 = \dl^e_{h_2h_1} \, ( L_{h_1} \, R^\ast_{h_1} L_{h_1^{-1}h_2h_1} - I ) \, ,
\ee
the triangle part ($h \in S_{(1)}$)
\be
   \R_{(h) \, h_1,h_2}
 = \dl^h_{h_2h_1} ( L_{h_1} \, R^\ast_{h_1} L_{h_1^{-1}h_2h_1} - L_h )
   \, {\cal E}_{(h)}
\ee
and the quadrangle curvature ($g \in S_{(2)}$)
\be
   \R_{(g) \, h_1,h_2;\hat{h}_1,\hat{h}_2}
 = \dl^g_{h_2h_1} ( L_{h_1} \, R^\ast_{h_1} L_{h_1^{-1}h_2h_1}
   - L_{\hat{h}_1} \, R^\ast_{\hat{h}_1} L_{\hat{h}_1^{-1} \hat{h}_2 \hat{h}_1} )
   \, {\cal E}_{(g)} \, .   \label{ocf_Rquadr}
\ee
Here we have introduced
\be
     {\cal E}^a_{(g) \, b}
  := \sum_{h' \in S} (R^\ast_g E^a{}_{h'}) \, \bar{E}^{g h' g^{-1}}{}_b \; .
\ee
\vskip.2cm

\noindent
{\it Example A.1.}
Let $G = \mathbb{Z}^2$ and $S = \{ \hat{1}=(1,0), \hat{2}=(0,1) \}$.
We choose a metric of Euclidean signature and a corresponding
orthonormal coframe $E^a$, $a = \mathbf{1},\mathbf{2}$, so that
\be
  \g_{h,h'} = \sum_{a,b} \delta_{a,b} \, E^a{}_h \, E^b{}_{h'} \; .
\ee
The metric-compatibility condition for a connection now reads
$L_h^T L_h = I$ and thus the matrices $L_h$ have to be orthogonal
$2 \times 2$-matrices. We may assume $\det L_h > 0$ so that
\be
  L_h(k,l) = \left(\begin{array}{cr}
   \cos \vartheta_h(k,l) & -\sin \vartheta_h(k,l) \\
   \sin \vartheta_h(k,l) & \cos \vartheta_h(k,l)
                   \end{array}\right)
\ee
for $(k,l) \in \mathbb{Z}^2$, which defines a map
$\mathbb{Z}^2 \times \{ \hat{1},\hat{2} \} \rightarrow SO(2)$.
The linear connection thus associates with each arrow on the
lattice a rotation angle.
Since there are no biangles or triangles in the case under
consideration, according to (\ref{ocf_Rquadr}) the curvature
is given by
\be
   \R_{\hat{1},\hat{2};\hat{2},\hat{1}}
 = ( L_{\hat{1}} \, R^\ast_{\hat{1}} L_{\hat{2}}
    - L_{\hat{2}} \, R^\ast_{\hat{2}} L_{\hat{1}} )
    \, {\cal E}_{(1,1)}
\ee
where ${\cal E}_{(1,1)} = (R^\ast_{(1,1)} E) \, \bar{E}$.
The last factor achieves that all indices of
the curvature tensor refer to the same point.

For the general metric (\ref{metric_abc}) and an arbitrary linear
connection, the curvature scalar can be expressed as follows,
\be
 R &=& \sum_{h,h_1,h_2 \in S}(\g^{-1})^{h_1,h_2} \, \R^h{}_{h_1,h,h_2}
   = \sum_{h \in S} \Big( (\g^{-1})^{h, \hat{2}} \,
     \R^{\hat{1}}{}_{h,\hat{1},\hat{2}}
     - (\g^{-1})^{h,\hat{1}} \, \R^{\hat{2}}{}_{h,\hat{1},\hat{2}} \Big)
              \nonumber \\
   &=& \sum_{h,h' \in S} (\g^{-1})^{h,[\hat{1}} (\g^{-1})^{\hat{2}],h'} \,
       \R_{h,h',\hat{1},\hat{2}}
    = \det(\g^{-1}) \, \R_{[\hat{1},\hat{2}],\hat{1},\hat{2}}
\ee
where we used the antisymmetry of the curvature tensor components
in the last two indices (which holds in the case under consideration),
the symmetry of the metric, and anti-symmetrization brackets.
With
\be
   \R_{[\mathbf{1},\mathbf{2}],\hat{1},\hat{2}}
 = \sum_{h,h' \in S} \bar{E}^h{}_{[\mathbf{1}} \, \bar{E}^{h'}{}_{\mathbf{2}]} \,
   \R_{h,h',\hat{1},\hat{2}}
 = (\det{\bar{E}}) \, \R_{[\hat{1},\hat{2}],\hat{1},\hat{2}}
\ee
and $(\det \bar{E})^2 = 1/\det\g$, we obtain the identity
\be
   \R_{[\mathbf{1},\mathbf{2}],\hat{1},\hat{2}}
 = R \, \sqrt{\det\g}
\ee
for the Einstein-Hilbert density.
\hfill $\blacksquare$

\end{appendix}


\begin{thebibliography}{99}
\addcontentsline{toc}{section}{\numberline{}References}
\bibitem{DMH02-gl} A.~Dimakis and F.~M\"uller-Hoissen, ``Differential
 geometry of group lattices,'' math-ph/0207014,
 to appear in {\em J. Math. Phys.} (2003).
\bibitem{Woro89} S.L. Woronowicz, ``Differential calculus on compact
 matrix pseudogroups (quantum groups),'' Commun. Math. Phys.
 {\bf 122}, 125 (1989).
\bibitem{BDMHS96} K.~Bresser, A.~Dimakis, F.~M\"uller-Hoissen and A.~Sitarz,
 ``Non-commutative geometry of finite groups,''
 J. Phys. A {\bf 29}, 2705 (1996).
\bibitem{Cast} L.~Castellani, ``Gravity on finite groups,''
 Commun. Math. Phys. {\bf 218}, 609 (2001);
 ``Finite group discretization of Yang-Mills and Einstein actions,''
 Ann. Phys. {\bf 297}, 295 (2002); \\
 P. Aschieri, L.~Castellani and A.P.~Isaev,
 ``Discretized Yang-Mills and Born-Infeld actions on finite
 group geometries,'' hep-th/0201223 (2002),
 ``Yang-Mills and Born-Infeld actions on finite group spaces,''
 hep-th/0210237; \\
 L. Castellani, R. Catenacci, M. Debernardi and C. Pagani,
 ``Noncommutative de Rham cohomology of finite groups,''
 math-ph/0211008.
\bibitem{Majid} F.~Ngakeu, S.~Majid and D.~Lambert,
 ``Noncommutative Riemannian geometry of the alternating group
 ${\cal A}_4$,'' J. Geom. Phys. {\bf 42} 259 (2002); \\
 S.~Majid, ``Riemannian geometry of quantum groups and finite groups
 with nonuniversal differentials,'' Commun. Math. Phys. {\bf 225}, 131
 (2002).
\bibitem{DMH99-dRg} A.~Dimakis and F.~M\"uller-Hoissen,
 ``Discrete Riemannian geometry,''
 J. Math. Phys. {\bf 40}, 1518 (1999);
 ``Pseudo-Riemannian metrics in models based on noncommutative geometry,''
 Czech. J. Phys. {\bf 50}, 45 (2000).
\bibitem{reg_graphs} Regular digraphs have constant valency (degree).
 There are regular graphs, like the well-known Peterson graph,
 which are not Cayley graphs. The differential calculus associated with
 a Cayley digraph has the distinguishing property that there is a
 (left) $\A$-module basis of the space of 1-forms. Moreover, the group
 structure organizes the arrows in a certain way which leads to the
 preferred basis $\{ \theta^h \, | \, h \in S \}$
 (see Ref.~\citen{DMH02-gl}).
\bibitem{DMH92} A.~Dimakis and F.~M\"uller-Hoissen, ``Noncommutative
 differential calculus, gauge theory and gravitation,'' report GOET-TP 33/92;
 ``A noncommutative differential calculus and its relation to
 gauge theory and gravitation,'' {\em Int. J. Mod. Phys. A (Proc.
 Suppl.)} {\bf 3A}, 474 (1993).
\bibitem{DGP02} C.~DiBartolo, R.~Gambini and J.~Pullin,
 ``Canonical quantization of constrained theories on discrete
 space-time lattices,'' {\em Class. Quantum Grav.} {\bf 19}, 5275 (2002); \\
 R.~Gambini and J.~Pullin, ``Canonical quantization of general relativity
 in discrete space-times,'' gr-qc/0206055,
 ``Discrete quantum gravity: applications to cosmology,'' gr-qc/0212033.
\bibitem{Regge61} T. Regge, ``General relativity without coordinates,''
 Nuovo Cim. A {\bf 19}, 558 (1961);
 R.M. Williams and P.A. Tuckey, ``Regge calculus: a brief review
 and bibliography,'' Class. Quantum Grav. {\bf 9}, 1409 (1992).
\bibitem{latt_grav}
 A. Das, M. Kaku and P.K. Townsend, ``Lattice
 formulation of general relativity,'' Phys. Lett. B {\bf 81}, 11  (1979); \\
 L. Smolin, ``Quantum gravity on a lattice,'' Nucl. Phys.
 B {\bf 148}, 333 (1979);  \\
 K.I. Macrea, ``Rotationally invariant field theory on lattices.
 III. Quantizing gravity by means of lattices,'' Phys. Rev. {\bf D23}
 900 (1981); \\
 C.L.T. Mannion and J.G. Taylor, ``General relativity on a
 flat lattice,'' Phys. Lett. B {\bf 100}, 261 (1981);   \\
 K. Kondo, ``Euclidean quantum gravity on a flat lattice,'' Progr. Theor. Phys.
 {\bf 72}, 841 (1984); \\
 G.~Friedberg, R.~Friedberg, T.D.~Lee and H.C.~Ren, ``Lattice gravity near
 the continuum limit,'' Nucl. Phys. B {\bf 245}, 145 (1984); \\
 H.~R\"omer and M.~Z\"ahringer, ``Functional integration and diffeomorphism
 group in Euclidean lattice quantum gravity,'' Class. Quantum Grav. {\bf 3},
 897 (1986); \\
 M. Caselle, A. D'Adda and L. Magnea, ``Lattice gravity and supergravity
 as spontaneously broken gauge theories of the (super) Poincar{\'e} group,''
 Phys. Lett. B {\bf 192}, 406 (1987), ``Doubling of all matter fields coupled to
 gravity on a lattice,'' Phys. Lett. B {\bf 192}, 411 (1987); \\
 P. Renteln and L. Smolin, ``A lattice approach to spinorial quantum gravity,''
 Class. Quantum Grav. {\bf 6}, 275 (1989); \\
 N.~Kawamoto and H.B.~Nielsen, ``Lattice gauge gravity,'' Phys. Rev. D {\bf 43},
 1150 (1991); \\
 O. Bostr\"om, M. Miller and L. Smolin, ``A new discretization of
 classical and quantum general relativity,'' preprint CGPG 94-3-3; \\
 L. Brewin, ``Riemann normal coordinates, smooth
 lattices and numerical relativity'', {\em Class. Quantum Grav.} {\bf 15}, 3085 (1998);
 ``An ADM 3+1 formulation for smooth lattice general relativity,''
 {\em Class. Quantum Grav.} {\bf 15}, 2427 (1998); \\
 R. Loll, ``Discrete approaches to quantum gravity in four dimensions,''
 {\em Living Rev. Rel.} {\bf 1}, 13 (1998); \\
 G.~Gionti, ``Discrete approaches towards the definition
 of a quantum theory of gravity,'' gr-qc/9812080.
\bibitem{Sauer70} R.~Sauer, {\em Differenzengeometrie} (Springer, Berlin, 1970).
\bibitem{rem:ambig}
The ambiguity in the quadrangle coefficient functions of a
2-form is not special to the $\cap$-product, but already appears in the
more general formalism using the original product in $\Omega$, see section~4.2
of Ref.~\citen{DMH02-gl}. In the latter work, $\check{\psi}_{h,h'}$ denoted
the quadrangle coefficients of a 2-form $\psi$ with respect to the
\emph{original} product in $\Omega$. Here it refers to the $\cap$-product.
Of course, they are different in general.
\bibitem{rem:conn_udc} Here the vanishing of the curvature is a special case
of a much more general well-known result. If the torsion of a linear
connection with respect to a differential calculus on some associative
algebra vanishes, then $\pi \nabla = \d \circ \pi$. In case of the
universal differential calculus this reduces to $\nabla = \d$ and the
vanishing of the curvature follows from $\d^2 = 0$.
\bibitem{Z4_math} A corresponding Mathematica notebook is available
 at the authors' homepages. Mathematica is a registered trademark of
 \emph{Wolfram Research}. See S.~Wolfram, {\em The Mathematica Book},
 (Cambridge University Press, 1999).
\bibitem{cnetwork} Distances in communication networks are measured by
sending a signal from one point to another and reflecting it back to
the point of origin. The elapsed time read off from
a clock at the point of emission, is then a measure of the distance
between the two points.
\bibitem{DFV91} M.~Dinneen, M.R.~Fellows and V.~Faber,
 ``Algebraic constructions of efficient broadcast networks,''
 in {\em Lecture Notes in Computer Science} {\bf 539}
 (Springer, 1991) p. 152.
\end{thebibliography}
\end{document}